\documentclass{iopart}
\usepackage{amssymb,mathdots}
\usepackage{iopams}
\usepackage{amsthm}
\usepackage{amscd}
\usepackage{amsfonts}
\usepackage[dvips]{graphicx}

\usepackage{cite}

\catcode`\@=11
\renewcommand\footnoterule{%
  \kern-3\p@
  \hrule\@width.4\columnwidth
  \kern2.6\p@}
\renewcommand\@makefntext[1]{%
    \parindent 1em\noindent
    \hb@xt@1.8em{\hss$^{\@thefnmark}$)}\hspace{2pt}%
    \footnotesize\rmfamily#1}  %\@makefnmark}#1}
\def\@makefnmark{\hspace{.5pt}\hbox{$^{\@thefnmark}$%
\hspace{-1pt}}} 

\newtheorem*{Th}{Theorem}

\newcommand{\ri}{{ \rm i }}

\newcommand{\nn}{\nonumber}

\newcommand{\be}[1]{\begin{equation}\label{#1}}
\newcommand{\ee}{\end{equation}}
\newcommand{\ba}[1]{\begin{eqnarray}\label{#1}}
\newcommand{\ea}{\end{eqnarray}}
\newcommand{\rf}[1]{(\ref{#1})}
\newcommand{\diag}{\mbox{\rm diag}\,}

\renewcommand{\Im}{\,\mbox{\rm Im}\,}
\renewcommand{\Re}{\,\mbox{\rm Re}\,}

\newcommand{\dd}{\dagger}
\newcommand{\mod}{\,\mbox{\rm mod}\,}

\def\RR{\mathbb{R}}

\def\CC{\mathbb{C}}

\def\ZZ{\mathbb{Z}}

\newcommand{\cH}{\mathcal{H}}

\newcommand{\cK}{\mathcal{K}}
\newcommand{\cM}{\mathcal{M}}
\newcommand{\cN}{\mathcal{N}}
\newcommand{\cP}{\mathcal{P}}
\newcommand{\cT}{\mathcal{T}}

\newcommand{\cU}{\mathcal{U}}
\newcommand{\cV}{\mathcal{V}}

\def\p{\partial}
\def\a{\alpha}

\def\g{\gamma}
\def\G{\Gamma}
\def\sg{\sigma}
\def\e{\epsilon}

\def\lb{\lambda}

\def\t{\theta}

\def\ra{\rangle}
\def\la{\langle}

\begin{document}
\title[$\cP\cT-$symmetric Bose-Hubbard model]{A non-Hermitian $\cP\cT-$symmetric Bose-Hubbard
model: eigenvalue rings from unfolding higher-order
exceptional points}
\author{E. M. Graefe$^{a}$, U. G\"unther$^{b}$,
H. J. Korsch$^a$, A. E. Niederle$^a$}
\address{$^a$ Technical University Kaiserslautern,\\ D-67663 Kaiserslautern, Germany\\
$^b$ Research Center Dresden-Rossendorf,  POB 510119,\\ D-01314
Dresden, Germany}
\eads{\mailto{graefe@physik.uni-kl.de},\quad \mailto{u.guenther@fzd.de},\quad
\mailto{korsch@physik.uni-kl.de},\quad \mailto{a.niederle@gmx.de}}
%\maketitle
\begin{abstract}
We study a non-Hermitian $\cP\cT-$symmetric generalization of an
$N$-particle, two-mode Bose-Hubbard system, modeling for example a
Bose-Einstein condensate in a double well potential coupled to a
continuum via a sink in one of the wells and a source in the other.
The effect of the interplay between the particle interaction and the
non-Hermiticity on characteristic features of the spectrum is
analyzed drawing special attention to the occurrence and unfolding
of exceptional points (EPs).  We find that for vanishing particle
interaction there are only two EPs of order $N+1$ which under
perturbation unfold either into $[(N+1)/2]$ eigenvalue pairs (and in
case of $N+1$ odd, into an additional zero-eigenvalue) or into
eigenvalue triplets (third-order eigenvalue rings) and $(N+1)\mod 3$
single eigenvalues, depending on the direction of the perturbation
in parameter space. This behavior is described analytically using
perturbational techniques. More general EP unfoldings into
eigenvalue rings up to $(N+1)$th order are indicated.
\end{abstract}\vspace{-3ex}
\submitto{\JPA} 
%\maketitle

\section{Introduction}
\label{sec_intro}

Physical models usually describe only a rather small separated
system uncoupled from the rest of the world. In quantum physics the
behavior of such a system is governed by a Hermitian Hamiltonian
operator. If one wants to take the coupling to some external world
into account, one ends up with the description of an open quantum
system. A somehow crude but instructive way to describe such open
quantum systems is the use of effective non-Hermitian Hamiltonians.
These descriptions in general yield complex eigenvalues whose
imaginary parts describe the rates with which an eigenstate decays
to the external world. Most often non-Hermitian Hamiltonians are
introduced heuristically, although this approximative description
can be achieved in a mathematically satisfactory way for example by
applying the Feshbach projection operator technique \cite{Okol03}.

Thinking of the description of open quantum systems it might be
surprising that there is a whole class of non-Hermitian Hamiltonians
which in some parameter regions give rise to purely real eigenvalues
and to unitary dynamics. These so-called $\cP\cT-$symmetric
Hamiltonians \cite{BB,BBjmp,PT-Z1,DDT-real,most-1,cmb-rev} possess
space-time-reflection symmetry, e.g., they commute with the $\cP\cT$
operator, where the operators $\cP$ and $\cT$ are defined by their
effects on the position and momentum operator $x$ and $p$
as
\ba{eq_PT}
\nn \cP: \qquad x \mapsto -x,\quad p\mapsto -p\\
\cT: \qquad x\mapsto \phantom{-}x,\quad p\mapsto
-p,\quad \ri\mapsto -\ri.
\ea
In some parameter region, the region of unbroken $\cP\cT-$symmetry,
all eigenvalues of $\cP\cT-$symmetric Hamiltonians are purely real
and the behavior of the system is similar to that of Hermitian
quantum systems. To get a feeling for the underlying reasons of this
"pseudo-closed" behavior of an open system, one can think of it in
terms of a balanced probability-flow \cite{Weig04}. Replacing the
condition of Hermiticity by the condition of $\cP\cT-$symmetry
therefore yields as well a fully consistent quantum theory, which
attracted a lot attention in the last years and actually stimulated
the research in other fields of physics like complexified classical
systems, supersymmetry and quantum field theory \cite{PT-ov}.

At first glance one might get the impression that the
non-Hermiticity is nothing but a small perturbation which does not
change the behavior of a system too much compared to the Hermitian
case, despite an additional decay behavior, or it may even be
equivalent to a Hermitian theory in the presence of
$\cP\cT-$symmetry. But actually non-Hermitian physics can differ
radically from Hermitian physics, especially in the presence of
eigenvalue degeneracies. While a Hermitian operator is always
diagonalizable (eigenvalues may coalesce, nevertheless they always
correspond to distinct eigenvectors), for a non-Hermitian
Hamiltonian the occurrence of nontrivial Jordan blocks in its
spectral decomposition is possible -- there may be points in
parameter space at which both eigenvalues {\it and} eigenvectors
coalesce, so-called exceptional points (EPs) \cite{kato,baumg}. The
occurrence of EPs in a system has drastic effects on the systems
behavior, especially concerning adiabatic features and geometric
phases. For the occurrence of EPs in various physical models see,
for example and not aiming at any completeness
\cite{mois-fried,heiss-1,mondragon-1,darmstadt,dorey-ep-2001,Okol03,berry-optics1,moiseyev2003,Korsch-Mossmann-1,oleg2004,berry-optics2,Seyr05,Guen05,Znoj07,GRS-EP-JPA,caliceti-2007,cejnar,quesne-ep-2007}.
In the theory of $\cP\cT-$symmetric quantum systems EPs naturally
occur as phase-transition points between sectors of exact
$\cP\cT-$symmetry and sectors of spontaneously broken
$\cP\cT-$symmetry.

The field of $\cP\cT-$symmetric Hamiltonians is still young and the
underlying mathematical structures are not completely understood
yet. Therefore in the last few years the interest in comparatively
simple systems with a finite dimensional Hilbert space, especially
$\cP\cT-$symmetric matrix Hamiltonians, was rapidly growing
\cite{Guen05,Weig06,Znoj07}. In this context most investigations
focused on the mathematical behavior of simple matrix models,
without demanding them to represent a physical system. Nevertheless
under special conditions there are a lot of physical systems which
indeed justify the description via a finite matrix model. In the
present paper we introduce a $\cP\cT-$symmetric generalization of a
prominent Hermitian matrix model, a two-mode Bose-Hubbard
Hamiltonian, which in the case of $N$ particles acts on an
$N+1$ dimensional Hilbert space.

The Bose-Hubbard Hamiltonian is a simple description of interacting
bosons on a lattice, which only takes one state per lattice site
into account. Originally the Hubbard model is a basic model of solid
state physics, where it is mostly addressed in its fermionic version
to describe the behavior of electrons in solids. In the last few
years it is enjoying a renaissance in the context of Bose-Einstein
condensates (BEC) in optical potentials. Due to the extremely low
temperatures and the precise periodicity of the optical potential
these systems provide the possibility of a clean experimental
realization of many kinds of theoretical models for interacting
many-particle systems. One prominent example is the
superfluid to Mott insulator phase transition which was realized
in a BEC in a three-dimensional optical potential \cite{Grei02}.

Investigating large $M$-mode, $N$-particle Bose-Hubbard systems quickly goes
beyond the scope of numerical manageability. Therefore, due to its
simple structure, the two-mode case, which one may think of
describing a BEC in a double well trap, became a
standard model \cite{Milb97, Angl01, Holt01a, Fran01, Mahm05, Wu06}. In
the present paper we introduce an effective non-Hermiticity to this
two-mode Bose-Hubbard Hamiltonian in a $\cP\cT-$symmetric way, which one
can imagine as an additional source and sink of equal strength. A
closely related -- and slightly more physical -- model would
include only a sink and would yield complex eigenvalues with a negative
imaginary part, describing a decay of particles. First theoretical
results for this non-Hermitian two-mode Bose-Hubbard system were
presented in \cite{Hill06}. As a possible realization one can think
of a BEC in a double well trap, where the condensate could escape
from one of the traps via tunneling. Another possibility would be
the outcoupling of atoms from one of the traps via radiofrequency
\cite{Bloc98}.

In the present paper we analyze the spectrum of the
$\cP\cT-$symmetric two-mode Bose-Hubbard system where we draw
special attention to the occurrence and the unfolding of EPs.
Numerical results are presented to illustrate the characteristic
behavior of this unfolding. Furthermore we use perturbative methods
which allow for analytic descriptions. Basic tools are the Le
Verrier-Faddeev method \cite{gantmacher} for the derivation of the
coefficients of characteristic polynomials of matrices and the
Newton-polygon technique for the extraction of the dominant powers
of polynomial perturbations \cite{baumg,chebotarev,trenogin}.

In detail the paper is organized as follows: In section
\ref{sec_model} we introduce the two-mode Bose-Hubbard Hamiltonian
and its $\cP\cT-$symmetric generalization and review the basic
vocabulary according to Hermitian and non-Hermitian ´degeneracies.
We discuss the analytically solvable limit of vanishing interaction
in section \ref{sec_lim_int} before we present numerical results on
the spectrum for non-vanishing interaction in section \ref{sec_num}.
Finally, in section \ref{sec_pert}, we investigate some of the
features of the spectrum previously found in the numerical studies
analytically using perturbative methods.

\section{Bose-Hubbard model and basic non-Hermitian vocabulary}
\label{sec_model}

The physical setup under consideration is a BEC in a double well
potential which at low temperatures can be analyzed in a two-mode
approximation. The corresponding Hamiltonian is that of a second
quantized many particle system of Bose-Hubbard type
\begin{equation} \label{Ham1}
  H = \varepsilon\left(a_1^{\dagger}a_1 - a_2^{\dagger}a_2\right) +
  v\left(a_1^{\dagger}a_2 + a_2^{\dagger}a_1\right) + \frac{c}{2}
  \left( a_1^{\dagger}a_1 - a_2^{\dagger}a_2\right)^2,
\end{equation}
where $a_j$, $a_j^\dagger$ are bosonic particle annihilation and
creation operators for the $j-$th mode, $2\varepsilon$ is the
on-site energy difference, $v$ controls the single particle
tunneling and $c$ the interaction strength between the particles.
In order to simplify the discussion we assume here that both $v$, and $c$ are
positive\footnote{Note that the energy spectrum stays the same if the sign of $v$ is altered, while it is turned upside down $E_n\to-E_n$ if the sign of $c$ is altered, which does not change the subsequent discussions in principle. Experimentally both negative and positive values of the interaction are possible and can actually be modulated via a Feshbach resonance \cite{feshbach}.}.
The Hamiltonian commutes with the particle number operator
\begin{equation} \label{Num}
N=a_1^{\dagger}a_1 +a_2^{\dagger}a_2,
\end{equation}
so that the total number $N$ of particles is conserved.

It is convenient to introduce angular momentum operators according to the Schwinger representation
\begin{eqnarray}\label{L}
  L_x &=& \frac{1}{2}\left(a_1^{\dagger}a_2 + a_2^{\dagger}a_1\right) \nonumber \\
  L_y &=& \frac{1}{2{\rm i}}\left(a_1^{\dagger}a_2 - a_2^{\dagger}a_1\right) \\
  L_z &=& \frac{1}{2}\left(a_1^{\dagger}a_1 - a_2^{\dagger}a_2\right),\nonumber
\end{eqnarray}
which obey the $su(2)$ commutation relation
 \be{comm}
 [L_x,L_y]={\rm i}L_z,
 \ee
and its cyclic permutations. In terms of these operators the
Hamiltonian (\ref{Ham1}) assumes the form
\begin{equation}\label{Ham_L}
  H = 2\varepsilon L_z + 2vL_x + 2cL_z^2.
\end{equation}
Thus for $\varepsilon,v,c\in\RR$ it is an element of the universal
enveloping algebra\footnote{For universal enveloping algebras see,
e.g., \cite{UEV-1,UEV-2,UEV-3}.} $\cU(su(2))$ of the $su(2)$ Lie
algebra in its angular momentum $l=N/2$ representation. In addition
we will often use the Lie algebra elements $L_\pm=L_x\pm iL_y$ with
commutation relations $[L_z,L_\pm]=\pm L_\pm$, \quad
$[L_-,L_+]=-2L_z$\,.

In the standard basis of the angular momentum algebra $|l,m\ra$, which can be defined
by the relations
\ba{ang-mom-4}\fl
L_\pm|l,m\ra =\sqrt{(l\mp m)(l\pm m+1)}|l,m\pm 1\ra, \qquad
L_z|l,m\ra=m|l,m\ra\,
\ea
with $l=N/2$, the Hamiltonian $H$ takes the form of a tridiagonal $(N+1)\times
(N+1)-$matrix
% \ba{eqn-ham-matrix}
% \langle j | H | k \rangle &=& H_ {j,k}, \qquad j,k=0,\ldots,
% N\nn\\
% H_{j,k} &=&  h_j(\varepsilon, c) \, \delta_{j,k} + v_j \,
% \delta_{j,k-1} + v_k\delta_{j-1,k}\nn\\
% h_j(\varepsilon, c)&:=& \varepsilon (2j-N) - \frac{c}{2} ( 2j
% -N)^2\nn\\
% v_j&:=& v \sqrt{(j+1) (N-j)}\,.
% \ea
\ba{eqn-ham-matrix} H=\left(
    \begin{array}{ccccc}
     d_l +c_l  & v_{l-1} &  \cdots & 0 &0 \\
      v_{l-1} & d_{l-1} +c_{l-1}&  \cdots & 0&0\\
      \vdots & \ddots &  \ddots & \vdots&\vdots\\
      0&0&\cdots & -d_{l-1} +c_{l-1}&v_{l-1}\\
      0 & 0  & \cdots & v_{l-1}&-d_l+c_l\\
    \end{array}
  \right)\nn\\
d_m:=2\varepsilon |m|,\qquad c_m:= 2cm^2,\quad -l\le m\le l\nn\\
v_m:=v\sqrt{(l+m+1)(l-m)}=v_{-(m+1)} \,.
\ea

In the following, for the non-Hermitian generalization of the
Bose-Hubbard Hamiltonian and the structure analysis of the
characteristic polynomials (see equation \rf{h-6} and below) another
representation of the angular momentum basis in terms of mononomials
in a complex variable will turn out to be most convenient. The
representation (\ref{ang-mom-4}) can be described in a standard way
as monomials in a variable $\xi\in\CC$ as (see, e.g.,
\cite{vilenkin,perelomov})
\ba{ang-mom-1}
|l,m\ra \cong f_m(\xi)=\frac{\xi^{l+m}}{\sqrt{(l-m)!(l+m)!}}\in \cH_l,
\qquad -l\le m\le l
\ea
with the normalization condition
\be{ang-mom-2}\fl
\la l,j|l,m\ra=\frac{(2l+1)2l!}\pi
\int\frac{\overline{f_j(\xi)}f_m(\xi)}{(1+|\xi|^2)^{2l+2}}d^2\xi=\delta_{jm},\quad
d^2\xi:=d(\Im\xi)d(\Re\xi) \,.
\ee
Here, $\cH_l$ denotes the space of polynomials in $\xi$ of degree
less or equal $2l+1$ \cite{vilenkin,perelomov}, in which the $SU(2)$
group representation acts. In the representation
\rf{ang-mom-1} the angular momentum operators act as first-order differential operators
\ba{ang-mom-3}
L_z=\xi\p_\xi-l,\qquad L_+=-\xi^2\p_\xi+2l\xi,\qquad L_-&=&\p_\xi
\ea
and yield again the relations (\ref{ang-mom-4}).

We note that (\ref{ang-mom-4}, \ref{ang-mom-1}, \ref{ang-mom-2}, \ref{ang-mom-3})
is the standard complex irreducible representation (irrep) of the real Lie algebra $su(2)$
(and the corresponding compact, simply connected group $SU(2)$) for
fixed angular momentum $l$ \cite{vilenkin,perelomov}. The advantage
of this complex $(2l+1)-$dimensional irrep is its straightforward
linear extendability to the complexification of $su(2)$, i.e., to
$sl(2,\CC)$ (see proposition 4.6 in \cite{hall-lie-rep}). Such a
complexification is necessarily encountered  when one passes from
real coefficients $\varepsilon$, $v$, $g$ in the Hamiltonian
\rf{Ham1} to complex ones --- as in our case when we pass from the
Hermitian $H$ to the non-Hermitian $\cP\cT-$symmetric $H$ by
assuming $\varepsilon$ to be purely imaginary. Under the
specific embedding $su(2)\hookrightarrow sl(2,\CC)$ the irrep
dimension $2l+1$ remains fixed. For completeness, we further note
that the complexification of the non-compact, not simply connected
real $SU(1,1)$ yields another embedding $su(1,1)\hookrightarrow
sl(2,\CC)$ with corresponding extensions of the infinite dimensional
$su(1,1)$ irreps \cite{perelomov,hall-lie-rep}. Below we will
consider boosts within the $(2l+1)-$dimensional irrep of the
$su(2)\hookrightarrow sl(2,\CC)$ embedding, not involving infinite
dimensional $su(1,1)-$related irreps (and corresponding matrices of
countably infinite order), i.e. we keep within the
$su(2)\hookrightarrow sl(2,\CC)$ induced irrep although boosts are
naturally connected with $SU(1,1)$ transformations.

The previous considerations justify the expansion of a non-Hermitian
generalization of the Hamiltonian \rf{Ham1} in the same
basis \rf{ang-mom-4} which yields a
non-Hermitian matrix representation as a generalization of the matrix \rf{eqn-ham-matrix}.

In the present paper we investigate a situation, where we assume the
on-site energy difference $\varepsilon$ of the two-mode Bose-Hubbard
model \rf{Ham1} to be complex, while the parameters $c$ and $v$ are
kept real. In particular, we focus on the case of a BEC in a
symmetric double well, where the real parts of the energies of both
modes are equal and therefore $\varepsilon$ is purely imaginary
$\varepsilon\equiv -i\g,\ \g\in\RR$
\be{pt-ham}
H = -2i\g L_z + 2vL_x + 2cL_z^2\,.
\ee
Physically such an imaginary on-site energy difference can be
achieved by coupling the modes to a continuum so that they will be
unstable --- decaying and amplifying in a balanced way. Although
$\varepsilon\in i\RR$ spoils the Hermiticity of the Hamiltonian $H$
in a usual Euclidian Hilbert space, it nevertheless leaves $H$
Hermitian in a Hilbert space with an indefinite inner product
structure, i.e., in a so-called Krein space
\cite{azizov,L2,japar,GSG,LT-1,Alb-Kuzhel-2004,GSZ-squire,Tanaka}.
This is easily seen from the explicit matrix structure, which is
essentially equivalent to \rf{eqn-ham-matrix} with the substitution
$\varepsilon\equiv -i\g$. The matrix $H$ is not only symmetric,
$H=H^T$, rather  it holds also
\be{pt-1}
H= \cP H^\dd \cP
\ee
where $\cP$ is the standard involutory permutation (sip) matrix
\be{sip}
\cP=\left(
      \begin{array}{ccccc}
        0 & 0 & \cdots & 0 & 1 \\
        0 & 0 & \cdots & 1 & 0 \\
        \vdots & \vdots & \iddots & \vdots & \vdots \\
        0 & 1 & \cdots & 0 & 0 \\
        1 & 0 & \cdots & 0 & 0 \\
      \end{array}
    \right),\qquad \cP^2=I
\ee
which is similar to an indefinite diagonal matrix
\be{sip2}\fl
\RR^{2n\times 2n}\ni \cP\sim \left(
                               \begin{array}{cc}
                                 I_n & 0 \\
                                 0 & -I_n \\
                               \end{array}
                             \right),\qquad \RR^{(2n+1)\times (2n+1)}\ni \cP\sim \left(
                               \begin{array}{cc}
                                 I_{n+1} & 0 \\
                                 0 & -I_n \\
                               \end{array}
                             \right).
\ee
Obviously, $\cP$ can be interpreted as parity operator which
interchanges the $a_1^\dd a_1-$ and $a_2^\dd a_2-$related modes in
\rf{Ham1} as
\be{sip3}
\cP:\quad |l,m\ra \ \mapsto \ \cP|l,m\ra=|l,-m\ra.
\ee
Denoting, as usual, the involution operator of the complex
conjugation -- the time reversal operator of quantum mechanics --
by $\cT$, where $\cT^2=I$, and taking into account that
\begin{equation}
H^\dd=\cT H^T\cT=\cT H\cT
\end{equation}
we find that the Hamiltonian $H$ for
$\varepsilon\in i\RR$ is $\cP\cT-$symmetric
\begin{equation}
[\cP\cT,H]=0.
\end{equation}
According to \rf{pt-1} it is self-adjoint in the Krein space
$\cK_{\cP}$ with the indefinite inner product $[.,.]_{\cP}=\la
.|\cP|.\ra$. Therefore the spectrum of $H$ will contain not only
real branches, but also pairwise complex conjugate branches and
exceptional points (branch points) at the transitions between real
and complex sectors of the spectrum. This is the typical behavior of
a $\cP\cT-$symmetric operator. For completeness, we note that purely
real branches correspond to parameter regions of exact
$\cP\cT-$symmetry ($H$ and its eigenfunctions are
$\cP\cT-$symmetric), whereas pairwise complex conjugate eigenvalues
correspond to regions of spontaneously broken $\cP\cT-$symmetry (in
contrast to $H$ its eigenfunctions are not $\cP\cT-$symmetric).

In the more general case of complex on-site energy difference
$\varepsilon\in\CC$ where neither real nor imaginary part are
vanishing, the $\cP\cT-$symmetry of the system is spoilt and one
obtains, in general, a spectrum not containing regions of purely
real eigenvalues.

For a Hermitian operator the spectrum is purely real. Possible level
crossing (degeneration) points will be so-called diabolical points
(DPs) \cite{berry-diab}, which are connected with diagonalizable
spectral decomposition, where algebraic multiplicity $n_a$ and
geometric multiplicity $n_g$ \cite{baumg} of the degenerate
eigenvalues $\lb$ coincide\footnote{Eigenvalues of diagonalizable
matrices are called semi-simple (see e.g. \cite{eom-semisimple}) and
in case of $n_g(\lb)=n_a(\lb)=1$ simple.}, $n_a(\lb)=n_g(\lb)$ and
one finds a symmetry enhancement\footnote{For a $k-$fold DP a $U(k)$
rotation symmetry occurs within the span of the degenerate
eigenvectors.}. In addition to these diabolic degeneration points,
for a non-Hermitian operator the occurrence of exceptional points
(EPs) is possible and even generic. EPs are parameter configurations
at which for a corresponding degenerate eigenvalue $\lb$ the
algebraic multiplicity $n_a(\lb)$ is exceeding the geometric
multiplicity, $n_g(\lb)<n_a(\lb)$. This is connected with a
non-diagonalizable spectral decomposition of the operator (matrix)
\cite{kato,baumg}, i.e., the formation of non-trivial Jordan-block
structures \cite{gantmacher,Seyranian-Mailyb-book} and Jordan chains
of algebraic eigenvectors (associated vectors). Subsequently, we use
the term $m$th-order EP for an EP which is associated with an
$m$th-order Jordan block in the spectral decomposition.

The EPs and DPs live on certain hypersurfaces $\cV_j$ in the
underlying parameter space $\cM\supset\cV_j$ of the model. They are,
in general, of various co-dimensions and form a so-called stratified
manifold $\cV=\bigcup_j \cV_j$ (see, e.g., \cite{stratification}).
Depending on a concrete parameter perturbation the system may move
along the stratified manifold $\cV$ passing from one degeneration
type to another one, or, more generically, it may escape from $\cV$
so that the degeneration disappears and an EP or DP unfolds into
non-degenerate eigenvalues.

In the following sections we will analyze the occurrence and the
unfolding of EPs for our Bose-Hubbard model. We will find that in
the limit of vanishing particle interaction the model can be solved
analytically and that there exist only two EPs of order $N+1$. In
dependence on the direction of the perturbation in parameter space
each of these EPs  unfolds either into $[(N+1)/2]$ eigenvalue
pairs\footnote{The notation $[a]$ stands for the floor function
which yields the highest integer less or equal $a\in\RR$.} (and, in
case of $N+1$ odd, into an additional zero-eigenvalue) or into
eigenvalue triplets (third-order eigenvalue rings) and $(N+1)\mod 3$
single eigenvalues.

\section{The limit of vanishing interaction\label{sec_lim_int}}

For vanishing interaction, $c=0$, the
Hamiltonian $H$ in \rf{pt-ham} is a complex linear combination of
$su(2)$ Lie algebra elements
\be{c0-1}
H=2(-i\g L_z+vL_x)\in sl(2,\CC).
\ee
\begin{figure}[htb]
\centering
\includegraphics[width=8cm]{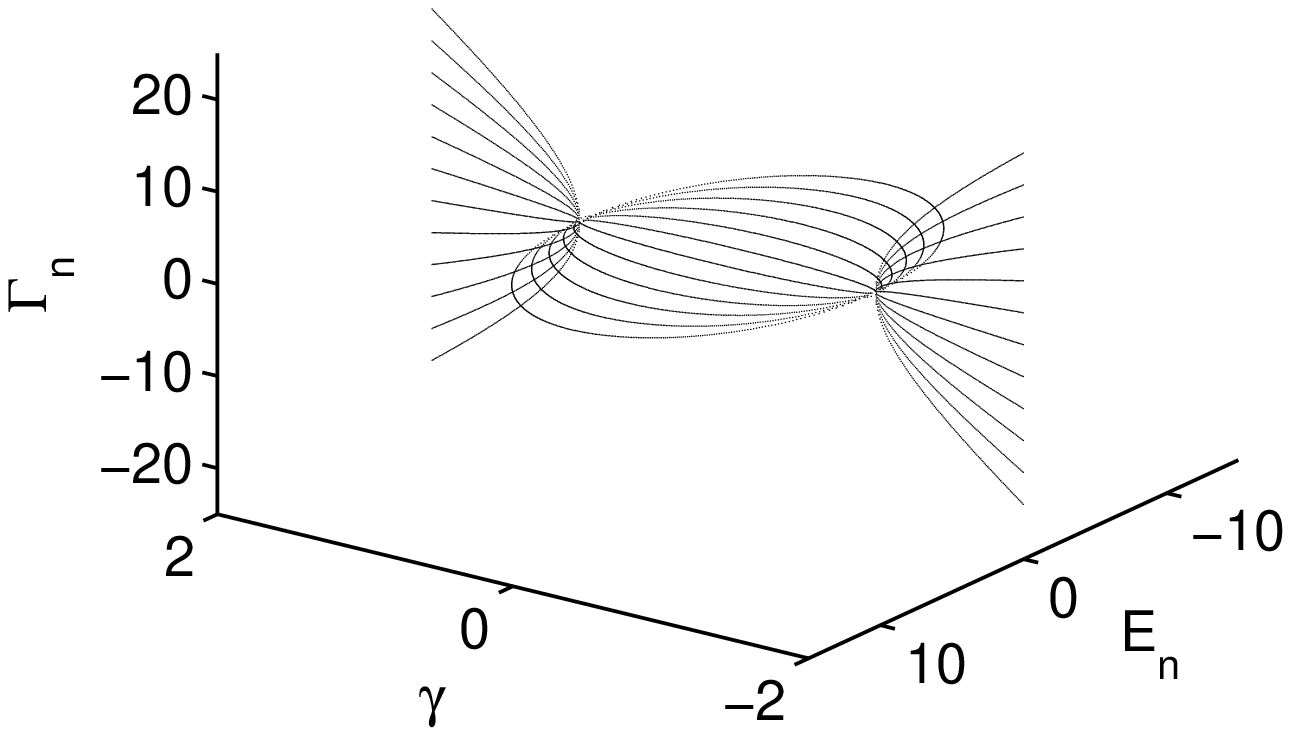}\\
\includegraphics[width=6.4cm]{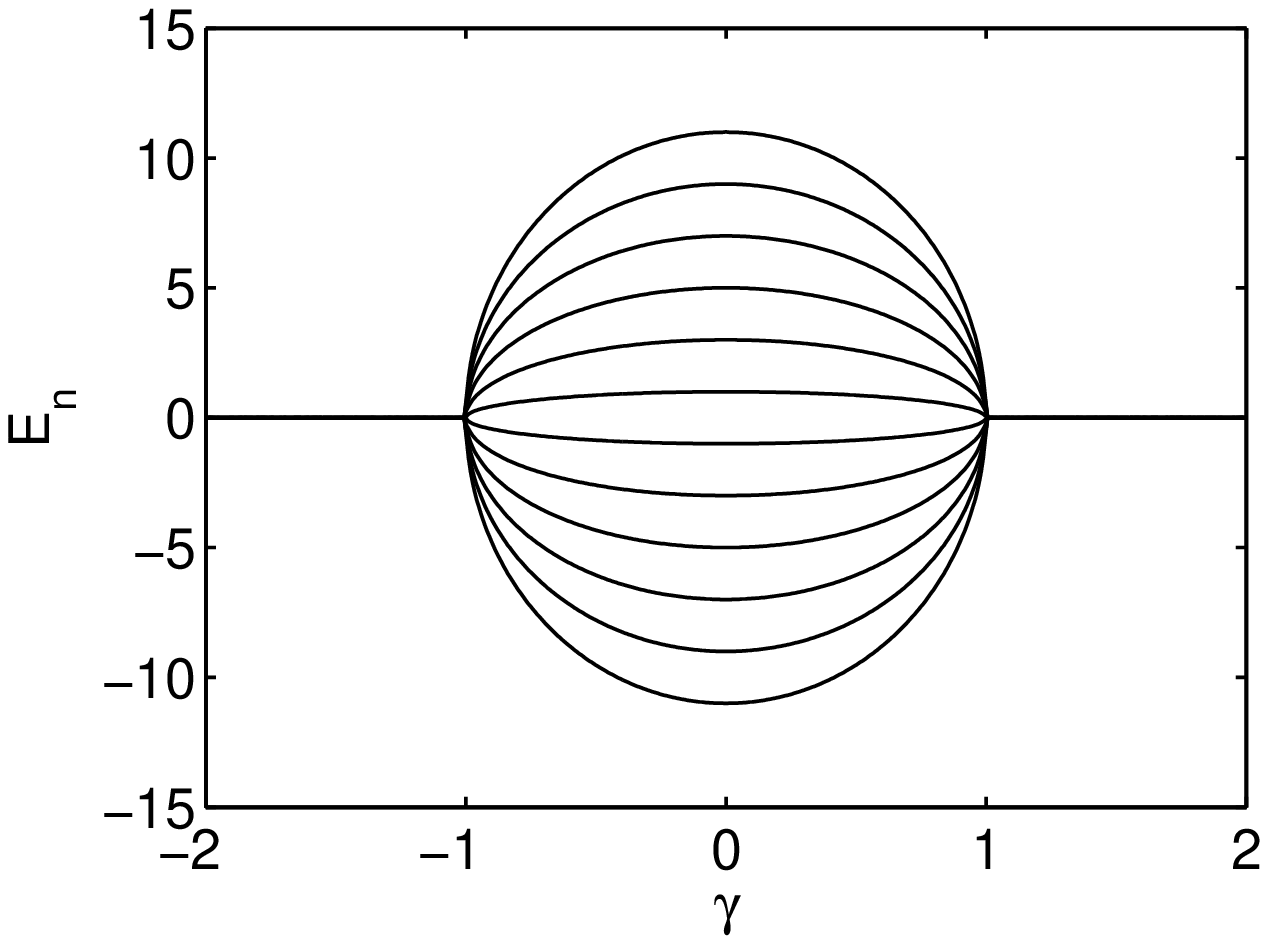}
\includegraphics[width=6.4cm]{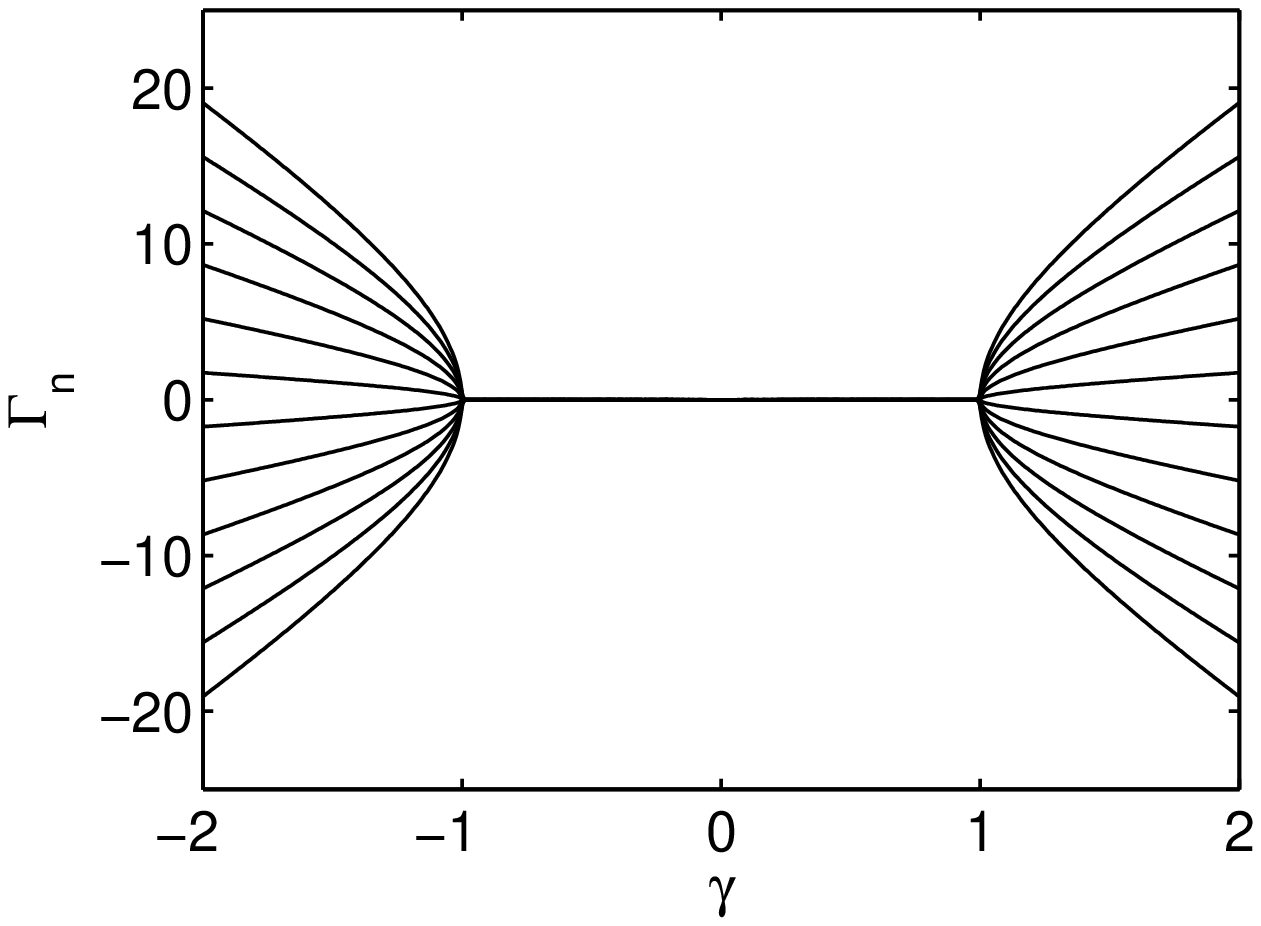}
\caption{\label{fig_expoint} Real- and imaginary parts of the eigenvalues $\lambda_n=E_n-\rmi\Gamma_n$ of the Bose-Hubbard Hamiltonian \rf{c0-1} as a function of the non-Hermiticity $\gamma$ for $v=1$ and $N=11$ particles for vanishing interaction $c=0$.}
\end{figure}
Figure \ref{fig_expoint} shows real and imaginary parts of the eigenvalues
of $H$ as a function of $\gamma$ for an example with $N=11$
particles and fixed $v=1$. All eigenvalues are purely real for
$|\g|<1$ and purely imaginary for $|\g|>1$. For $|\g|=1$ we observe
a degeneracy of all eigenvalues, which will turn out to correspond
to a full Jordan block, resp. an EP of order N+1. Furthermore the
real and imaginary parts are axis symmetric with regard to both axes
$E_n=0$ and $\G_n=0$
--- a behavior which is due to the high symmetry of the Hamiltonian \rf{c0-1}.
The eigenvalues can be easily obtained analytically by diagonalizing
$H$ in case of $|\g|\neq |v|$ and bringing it to its non-trivial
Jordan block form in case of $|\g|=|v|$.

For this purpose we make use of the Baker-Campbell-Hausdorff formula
and the $su(2)$ commutation relations \rf{comm} to obtain the well
known rotations and boosts over the algebra $su(2)$ specifically
needed for our analysis
\ba{c0-2}
e^{-i\t L_y}L_ze^{i\t L_y}&=&\cos(\t) L_z+\sin(\t)
L_x,\label{rot-y}\\
e^{\a L_y}L_ze^{-\a L_y}&=&\cosh(\a) L_z+i\sinh(\a)
L_x,\label{boost-y}\\
e^{-i\t L_x}L_ye^{i\t L_x}&=&\cos(\t) L_y +\sin(\t)
L_z\label{rot-x}\,.
\ea
With the help of the boost \rf{boost-y} and the identification
\be{c0-3}
\cosh(\a)=\frac \g{\sqrt{\g^2-v^2}}\,,\qquad \sinh(\a)=\frac
v{\sqrt{\g^2-v^2}}
\ee
the Hamiltonian \rf{c0-1} can be reshaped as
\ba{c0-4}
H&=&-2i\sqrt{\g^2-v^2}\,\left[\cosh(\a)L_z+i\sinh(\a)L_x\right]\nn\\
&=&-2\sqrt{v^2-\g^2}e^{\a L_y}L_ze^{-\a L_y}\,.
\ea
{}From the fact that the irrep \rf{ang-mom-1} remains valid for any
complex extension of $su(2)$ it follows the completeness of the
basis vectors
\be{c0-5}
I=\sum_{m=-l}^l |l,m\ra\la l,m|
\ee
and with it
\ba{c0-6}\fl
\la l,j|H|l,k\ra&=&-2\sqrt{v^2-\g^2}\sum_{m,m'=-l}^l \la l,j|e^{\a
L_y}|l,m\ra\la l,m|L_z|l,m'\ra\la l,m'|e^{-\a
L_y}|l,k\ra\nn\\
\fl &=&-2\sqrt{v^2-\g^2}\sum_{m=-l}^l S_{jm}^{-1}mS_{mk}
\ea
where
\be{c0-6b}
S_{mk}:=\la l,m|e^{-\a L_y}|l,k\ra.
\ee
For $|\g|\neq |v|$ the boost functions \rf{c0-3} remain finite so
that $|\a|<\infty$ and $S$ remains regular.  Therefore, $S$ acts as
similarity transformation which diagonalizes the Hamiltonian
\rf{c0-1}. The eigenvalues of $H$ can be read off from \rf{c0-6} as
\ba{Ev_lin}
\lambda_n=E_n-i\Gamma_n=n\sqrt{v^2-\gamma^2}
\ea
where $n=-N,-N+2,\dots,N-2,N$. In the limit $\g\to \pm v$  the
diagonalization breaks down and the boost becomes singular:
$|\a|\to\infty$. Instead of eq. \rf{boost-y} we may use a rotation
\rf{rot-x} with $\t=-\pi/2$, i.e.
\be{c0-6a}
e^{i\pi L_x/2}L_ye^{-i\pi L_x/2}=-L_z\,.
\ee
This allows us to represent the Hamiltonian as
\ba{c0-7}
H=2v(\mp i L_z+L_x)&=&2ve^{i\pi L_x/2}(\pm iL_y+L_x)e^{-i\pi
L_x/2}\nn\\
&=&2ve^{i\pi L_x/2}L_\pm e^{-i\pi L_x/2},
\ea
so that with $R_{nk}:=\la l, n|e^{-i\pi L_x/2}|l, k\ra$ we have
\be{c0-8}
\la l, j|H|l, k\ra=2v R^{-1}_{jm}\la l, m|L_\pm |l, n\ra R_{nk}\,, \qquad
\det(R)\neq 0.
\ee
According to \rf{ang-mom-4} only the elements $\la l, k\pm 1|L_\pm
|l, k\ra$ are non-vanishing. This means that the matrix $\la l, m|L_\pm
|l, n\ra$ is subdiagonal for $L_-$ and superdiagonal for $L_+$.
Matrices with all elements $\{a_n\}_{n=1}^N$ on their sub- or
superdiagonals non-vanishing and all other elements equal to zero
are similar to a single Jordan block $J_{N+1}(0)$ with eigenvalue
$\lb=0$:
\ba{c0-9}
\left(
  \begin{array}{cccccc}
    0 & a_1 & 0 & \cdots &0 & 0\\
    0 & 0 & a_2 & \cdots & 0 & 0\\
    \vdots & \vdots & \vdots& \ddots & \vdots &  \vdots\\
    0 & 0 & 0 & \cdots & a_{N-1} & 0\\
    0 & 0 & 0 & \cdots & 0 & a_N\\
    0 & 0 & 0 & \cdots & 0 & 0\\
  \end{array}
\right)=Q^{-1} J_{N+1}(0)Q,\nn\\
Q=\diag\left(1,a_1,a_1a_2,\cdots,\prod_{k=1}^Na_k\right).
\ea
This is also immediately evident from the angular momentum
operators: The ladder operators $L_\pm$ should only have a single
eigenstate, namely "spin-up" resp. "spin-down", belonging to the
eigenvalue $\lb=0$. Since for $|\g|=|v|$ the Hamiltonian is
equivalent to $L_\pm$, with total angular momentum $N/2$, these
configurations correspond to a degenerate eigenvalue $\lb=0$ which
is an EP of order $N+1$. Under variation of $\gamma$ it unfolds into
$[(N+1)/2]$ eigenvalue pairs (and in case of $N+1$ odd, into an
additional zero-eigenvalue) according to \rf{Ev_lin}. The result is
illustrated in figure \ref{fig_expoint}. We note that although such
a pairwise unfolding of an $(N+1)$th-order EP according to a square
root law $\sqrt{v^2-\gamma^2}$ with different scaling pre-factors
$n$ appears physically rather generic, it is mathematically very
special. Typically an $(N+1)$th-order EP at $\lb=0$ unfolds under a
small perturbation $|\e|\ll 1$ into an $(N+1)-$ring $\lb\approx
e^{i\frac{2\pi k}{N+1}}\e^{1/(N+1)}$, \ $k=0,\ldots, N$ of
non-degenerate eigenvalues. This follows simply from an effective
equation of the type $\lb^{N+1}\approx \e$. Intuitively, one might
expect the spectral behavior \rf{Ev_lin} with its decoupled square
roots to result from a decomposition of $H$ into $[\frac{N+1}2]$
second-order Jordan blocks rather than from a single $(N+1)$th-order
Jordan block. The subject of the remaining sections will be to
further clarify this specific (mathematically non-standard) spectral
behavior of $H$.

In contrast to the simple analytical structure of the model and
its complete solvability in case of vanishing interaction $c$, the
situation changes drastically for non-vanishing interaction. In this
case the spectrum of the non-Hermitian $\cP\cT-$symmetric
Hamiltonian \rf{pt-ham} is most efficiently studied with the help of
numerical and perturbational techniques. In the next section we will
present some numerical results for non-vanishing interaction.
Qualitative aspects of these findings are explained
analytically with the help of perturbational techniques in section
\ref{sec_pert}.

\section{Numerical results for non-vanishing interaction}
\label{sec_num}

\begin{figure}[htb]
\centering
\includegraphics[width=8cm]{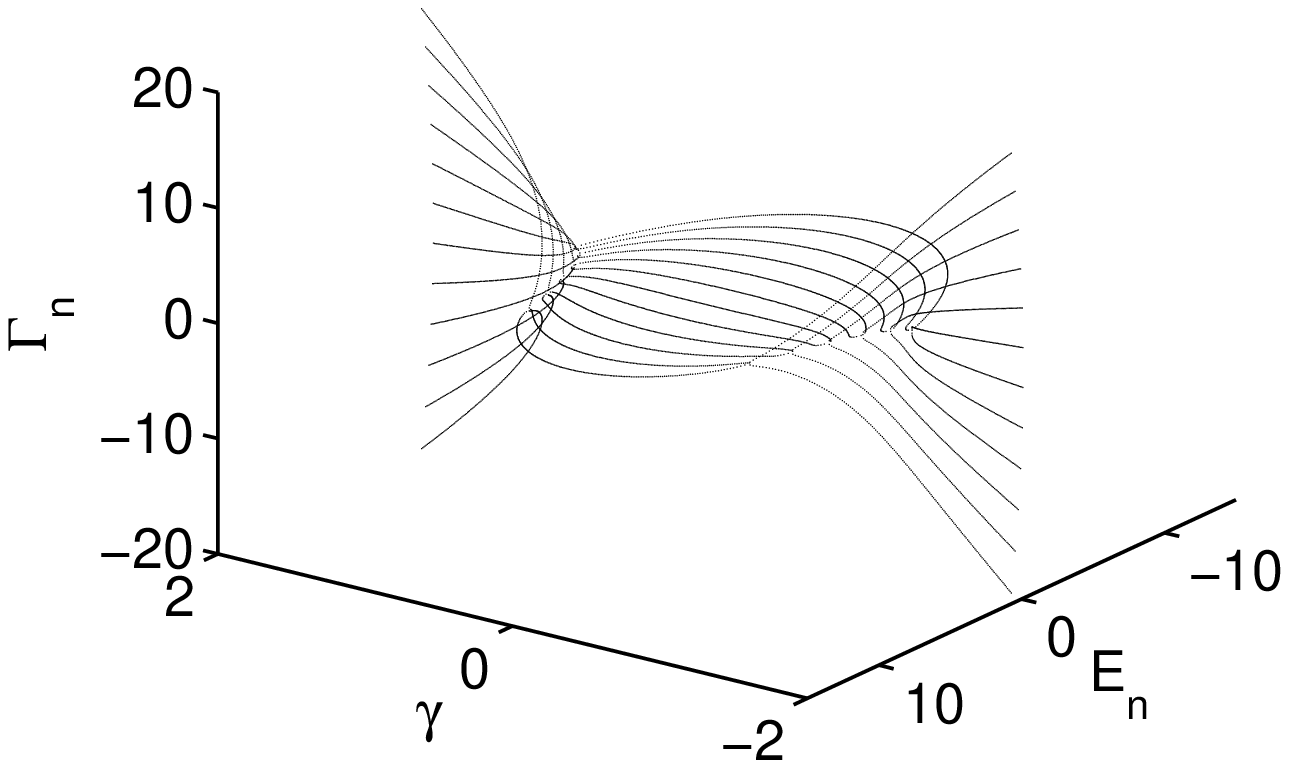}\\
\includegraphics[width=6.4cm]{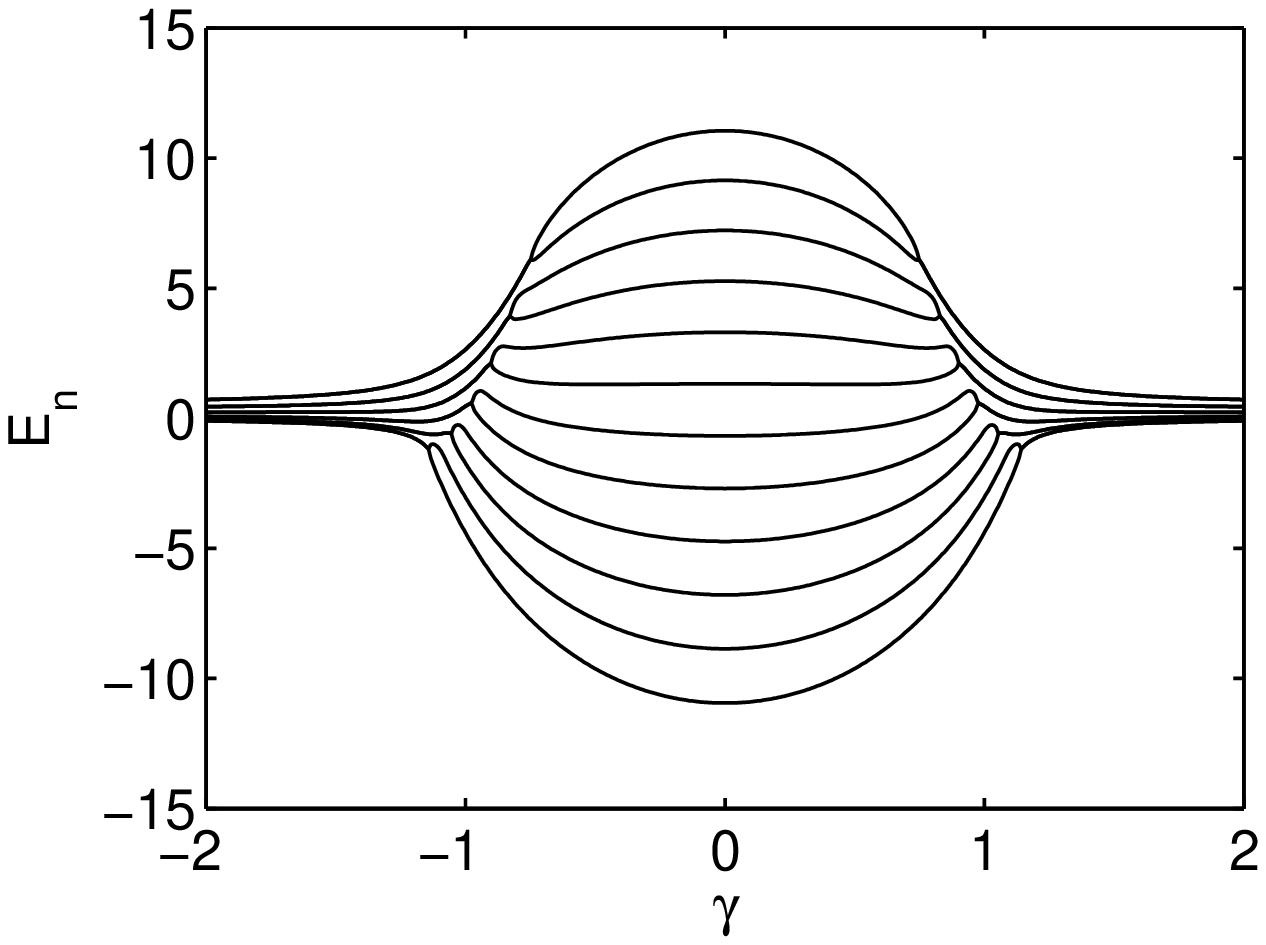}
\includegraphics[width=6.4cm]{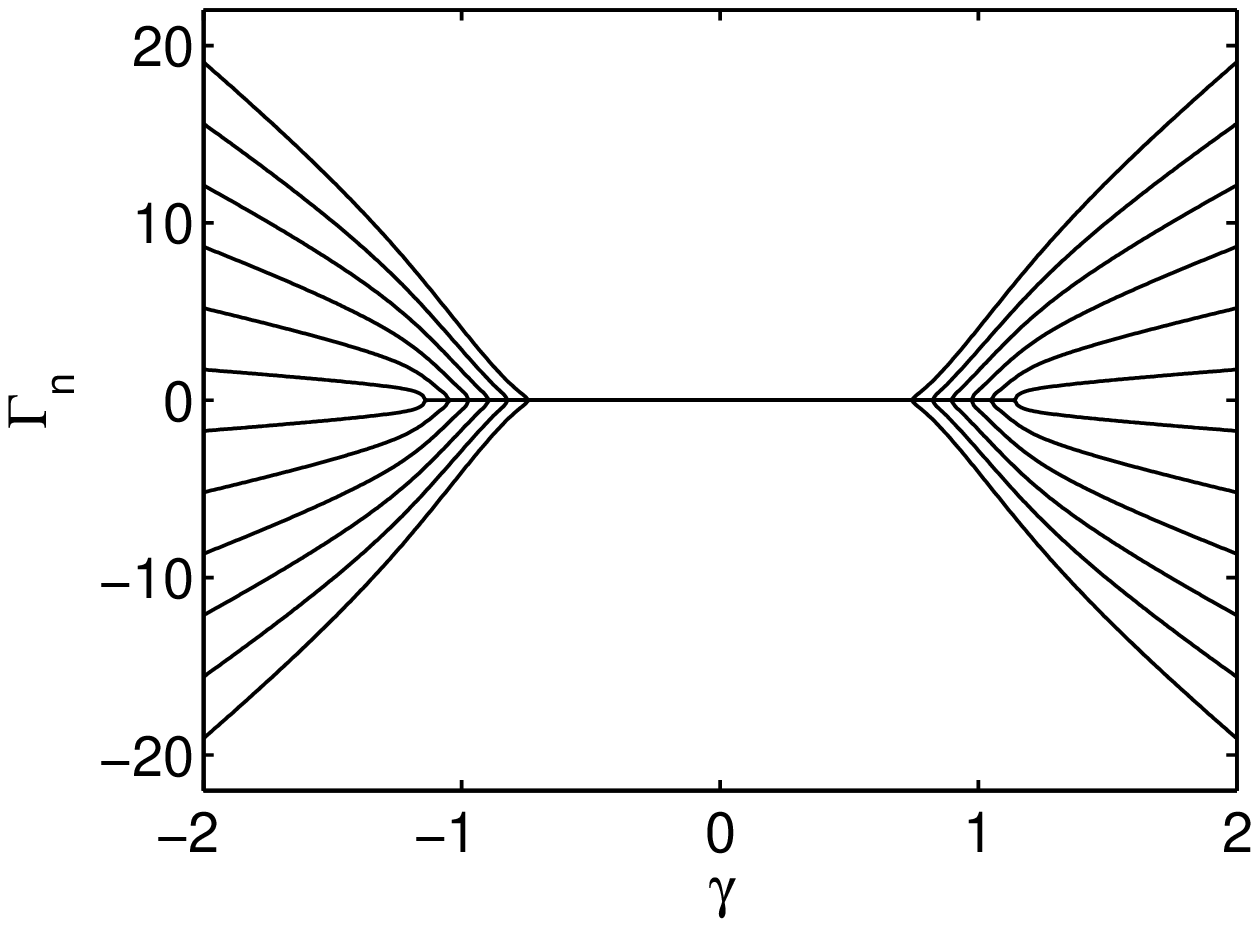}
\caption{\label{fig_EWgamma_c01} Real- and imaginary parts of the
eigenvalues $\lambda_n=E_n-\rmi\Gamma_n$ of the Bose-Hubbard Hamiltonian \rf{pt-ham} as a function of the non-Hermiticity
$\gamma$ for $v=1$, $c=0.1/N$ and $N=11$ particles.}
\end{figure}

\begin{figure}[htb]
\centering
\includegraphics[width=8cm]{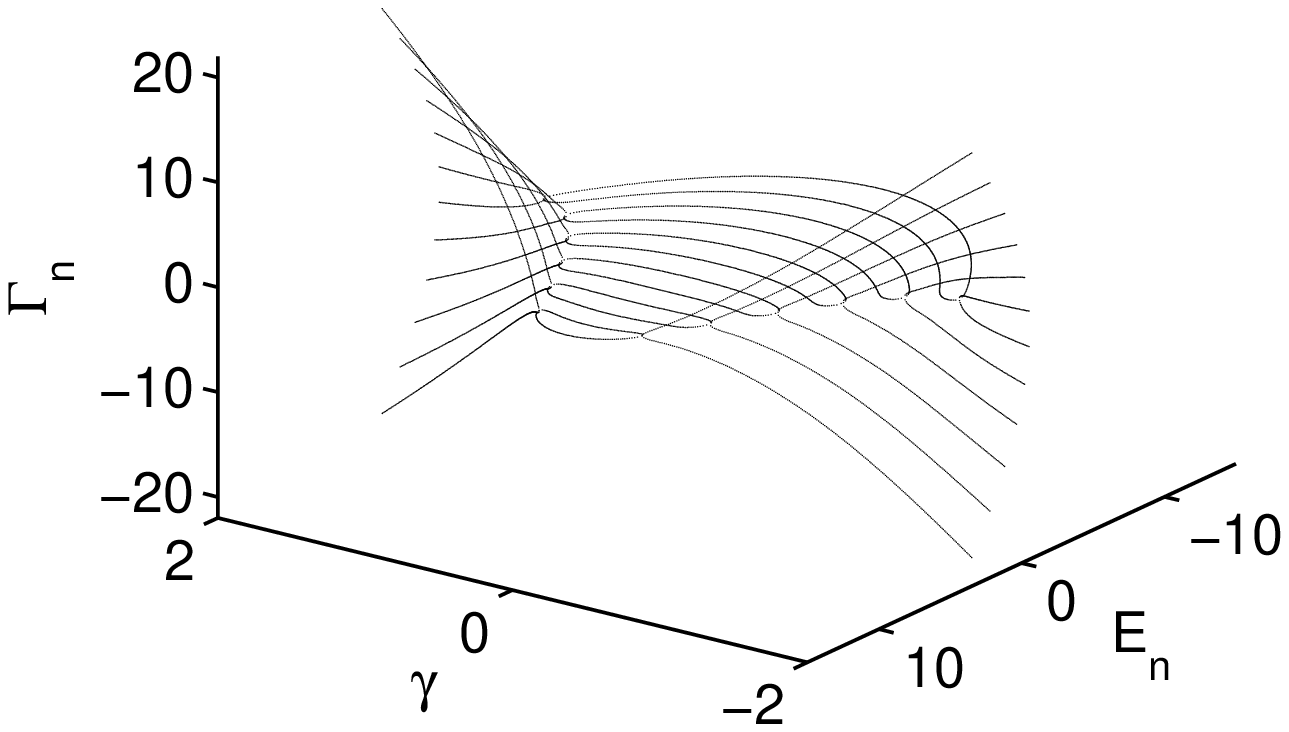}\\
\includegraphics[width=6.4cm]{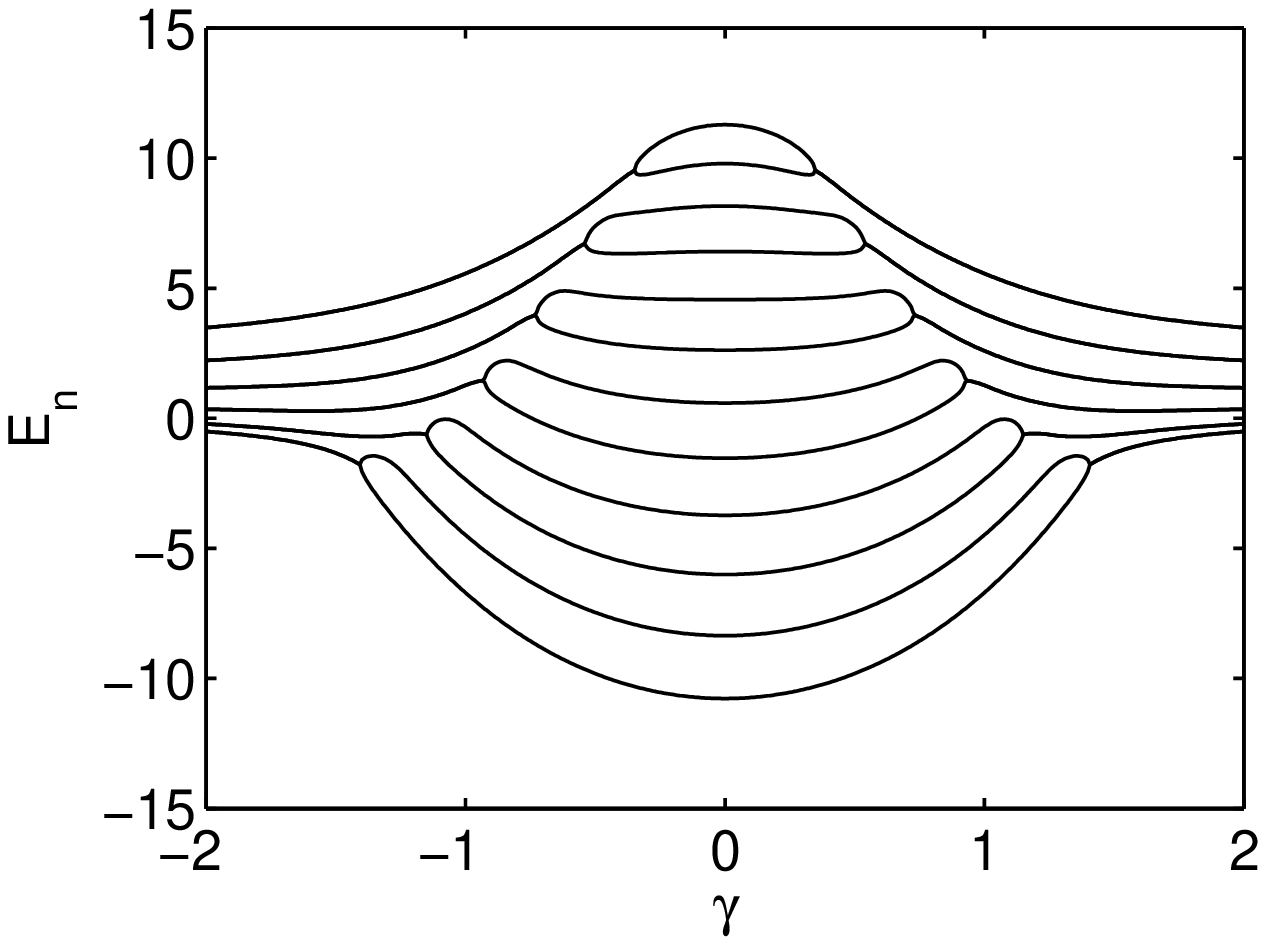}
\includegraphics[width=6.4cm]{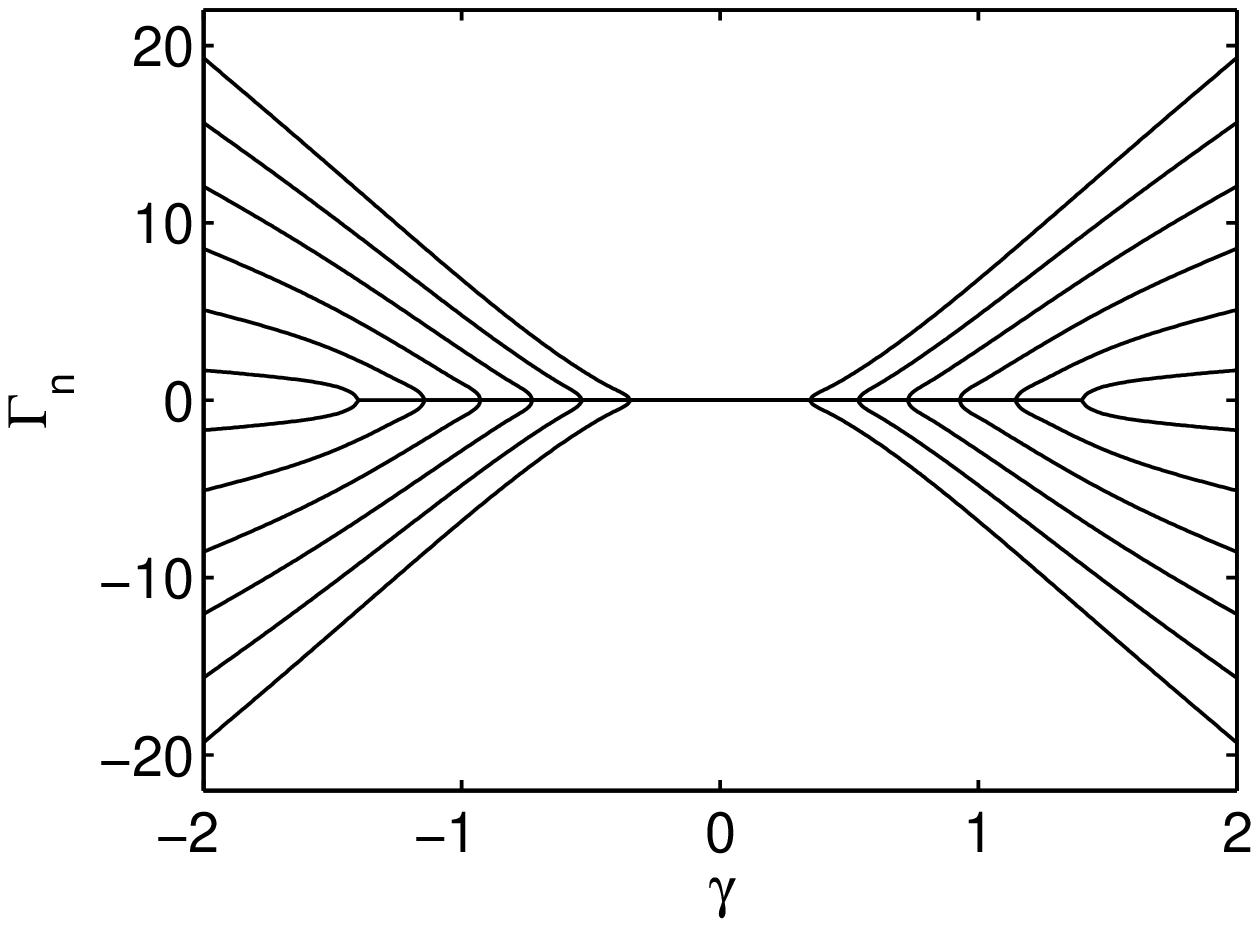}
\caption{\label{fig_EWgamma_c05} Real- and imaginary parts of the
eigenvalues $\lambda_n=E_n-\rmi\Gamma_n$ of the Bose-Hubbard Hamiltonian \rf{pt-ham} as a function of the non-Hermiticity
$\gamma$ for $v=1$, $c=0.5/N$ and $N=11$ particles.}
\end{figure}

\begin{figure}[!htb]
\begin{center}
  \includegraphics[width=6cm]{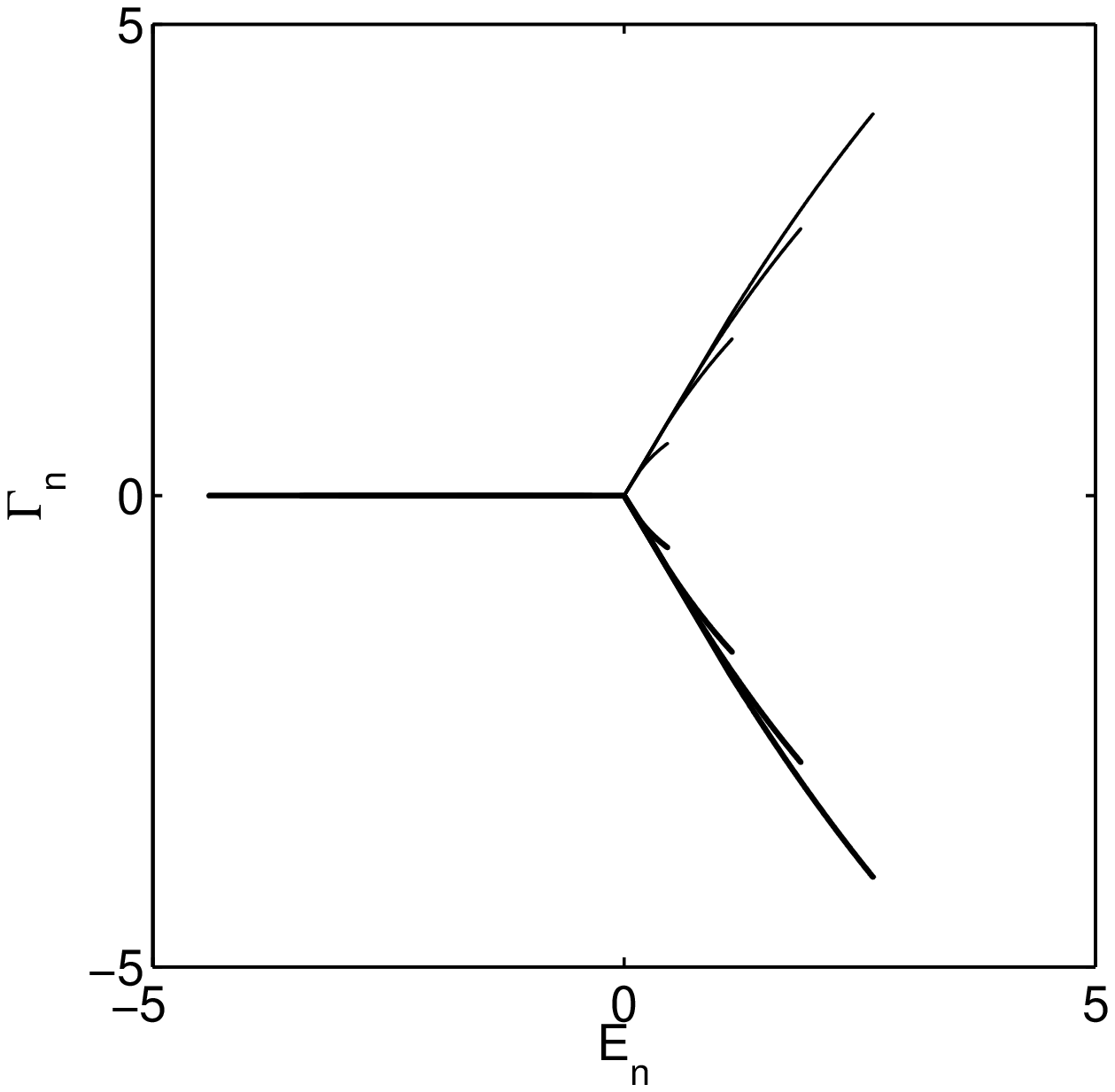}
  \includegraphics[width=6cm]{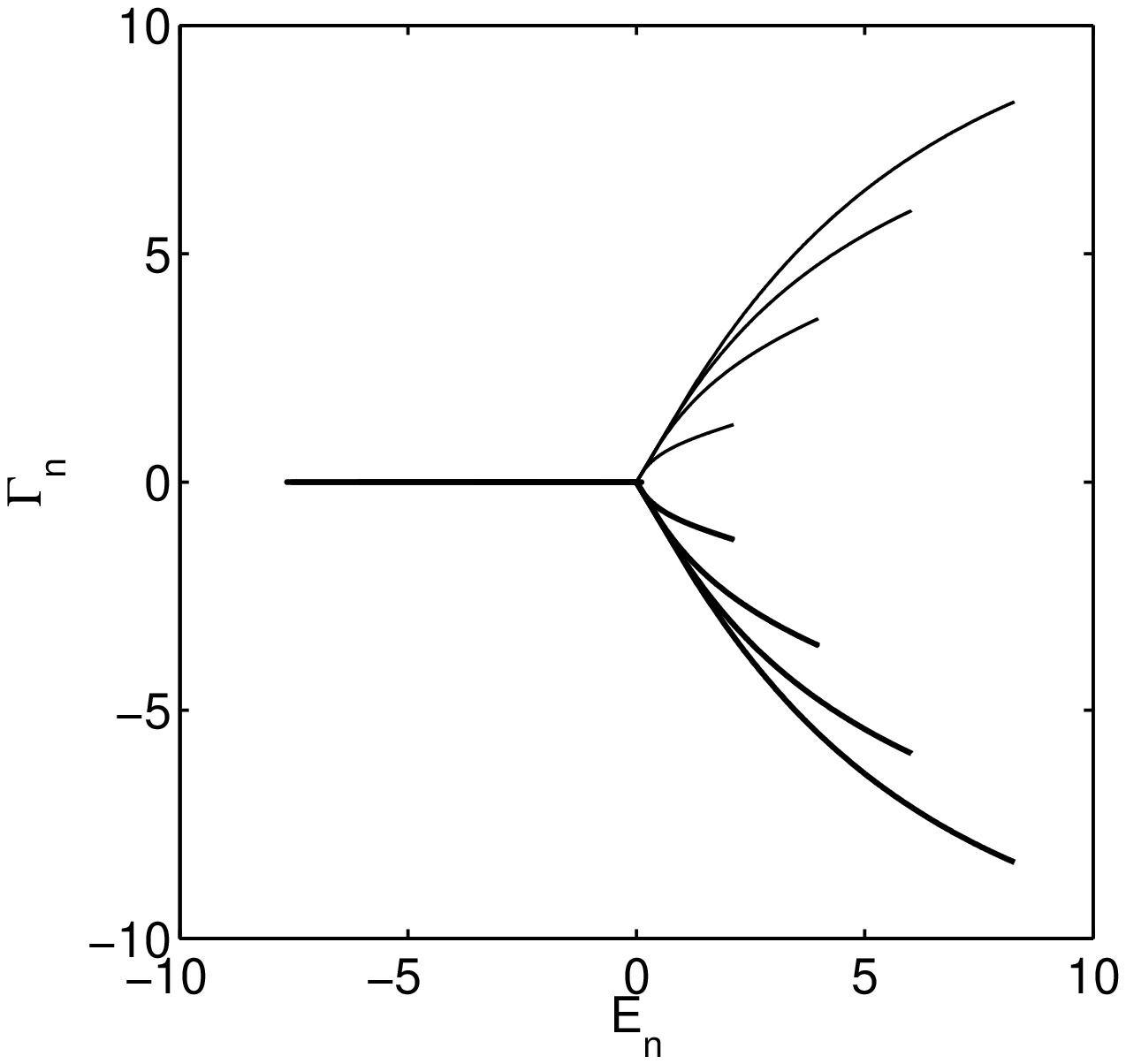}
 \caption{\label{fig_EVtrajectory11} Trajectories of the complex energy eigenvalues $\lambda_n=E_n-\rmi\Gamma_n$ of the Bose-Hubbard Hamiltonian \rf{pt-ham} as a function of $c$ with $0<cN<0.1$ (left) and $0<cN<1$ (right), for $\gamma=v$, $N=11$ particles.}
\end{center}
\end{figure}

Figures \ref{fig_EWgamma_c01} and \ref{fig_EWgamma_c05} demonstrate
the typical spectral behavior of a Hamiltonian \rf{pt-ham} in
dependence on the non-Hermiticity parameter $\gamma$ for fixed weak
and moderate interaction strength $c$. In the concrete example, the
twelve eigenvalues of an $N=11$ particle system are shown.

First we note that the non-vanishing interaction $c$ reduces the
symmetry of the spectrum: The symmetries with regard to sign changes
of the complex coupling $\g\rightleftharpoons -\g$ and of the
imaginary spectral components $\G_n\rightleftharpoons -\G_n$ are not
altered. The first symmetry results from the fact that, regardless
of the interaction, for symmetric modes it does not matter which of
them is coupled to the source and which to the sink. Due to this
$\g\rightleftharpoons -\g$ symmetry we can restrict our analysis to
the parameter region $\g\ge 0$. The second symmetry (with regard to
$\G_n\rightleftharpoons -\G_n$) is a direct implication of the Krein
space symmetry \rf{pt-1} of $H$ which causes non-real eigenvalues to
occur always in complex conjugate pairs. The additional symmetry
$E_n \rightleftharpoons -E_n$ present for $c=0$ is lost in case
$c\neq 0$. This is due to the fact that for $c\neq 0$ the square
root branch points (EPs) become shifted in different ways and do no
longer coalesce in the parameter space $\cM\cong \RR^3\ni (\g,v,c)$.
For $|c|\ll |v|/N$, $\g^2\approx v^2$ this behavior can be roughly
described as an effective deformation of the spectral branches
\rf{Ev_lin} of the type
\be{c0-10}\fl
\lb_n(c)=a_n(c)+b_n(c)n\sqrt{v^2-\g^2-d_n(c)}\qquad
a_n(0)=d_n(0)=0,\quad b_n(0)=1
\ee
which for $c\neq 0$ leads to
\be{c0-11}
E_n=\Re[\lb_n(c)]\neq -E_{-n}=-\Re[\lb_{-n}(c)]\,.
\ee
{}From figures \ref{fig_EWgamma_c01} and \ref{fig_EWgamma_c05} one
clearly sees that for fixed $v=1$ each of the two EPs of order
twelve present in figure \ref{fig_expoint} at $\g=\pm v$ splits up
into six second-order EPs with different positions $\g_n =\pm
\sqrt{v^2-d_n(c)}$. The fact that these special points are EPs is
obvious from the graphics. One clearly sees, firstly, that these
points are associated with transitions from real spectral branches
to complex conjugate ones and that they are therefore branch points.
Secondly, one observes that the lines which branch off from these
points scale faster than linearly so that the points cannot be
diabolical points which are connected with a linear scaling (see
e.g. \cite{Seyr05}).

For increasing $|c|$ the deviations from the spectrum at $c=0$ also
increase (compare figures \ref{fig_expoint}, \ref{fig_EWgamma_c01}
and \ref{fig_EWgamma_c05}).

Subsequently, it is convenient to relabel the absolute values
of the EP positions $|\g_n|=:\tilde \g_{k=\sg(n)}$ in increasing
order as
\be{c0-10a}
0\le \tilde \g_1\le \tilde \g_2\le \cdots .
\ee
Furthermore, we dub the $(N+1)$th-order EPs as mother EPs.

Complementary information about the unfolding of these mother EPs
for non-vanishing interaction strength $c\neq 0$ can be gained by
considering the behavior of the spectral branches at the former EP
positions $|\g|=|v|$. For this purpose the trajectories of the
twelve eigenvalues in the complex plane have been plotted for
interaction strengths $c$ varying in the intervals $0< c\le c_{\rm
max}=0.1/N $ and $0< c\le c_{\rm max}=1/N $ (see figure
\ref{fig_EVtrajectory11}). For definiteness we have chosen $v=1$ so
that the mother EP is localized at $\g=v=1$. Obviously, for small
$|c|\ll |v|/N $ the twelve eigenvalues form three groups where four
eigenvalues behave qualitatively almost identical
--- moving along one of the three lines in the directions $\sim
e^{-i\pi}, e^{-i\pi \pm i\frac{2\pi}3}$. This regular circle
division with lines enclosing angles of $2\pi/3$ in the complex
plane clearly indicates on an unfolding of the type
\be{co-12}\fl
\lb_{j,k}\sim (-f_j)^{1/3} e^{i\frac{2\pi k}3}c^{1/3}, \qquad
k=0,1,2,\qquad f_j\in \RR^+, j=1,2,3,4
\ee
where three eigenvalues corresponding to the same $f_j$ can be
understood as a triplet. For larger values of $c$ (see , e.g., the
right side of figure \ref{fig_EVtrajectory11}) one eigenvalue of
each triplet stays on the negative real axis, while the other two
depart from their initial directions symmetrically with regard to
the real axis as complex conjugate pairs.

The triplet splitting is less obvious from the spectral branches
depicted in figures \ref{fig_EWgamma_c01} and
\ref{fig_EWgamma_c05}. Nevertheless, these plots also clearly show
that at the position $\g=v=1$  four purely real eigenvalues and four
complex conjugate pairs of eigenvalues are present. Obviously, the
triplet splitting is closely connected with the fact that for
small $c\neq 0$ two of the second-order branch points (EPs) move
to values $\tilde \g_5,\tilde \g_6 >1$ yielding in this way four
real eigenvalues at $\g=1$, whereas the other four second-order
EPs with $\tilde \g_{1},\ldots,\tilde \g_4$ move to positions
$0<\tilde \g_k<1$, what results in the four complex conjugate pairs
at $\g=1$.

Further numerical investigation shows that the triplet unfolding of
the mother EP for $c\neq 0$ and $|\g|=|v|$ is a generic feature of
the system \rf{pt-ham} for arbitrary particle number. In general, we
find a splitting into $\left[\frac{N+1}{3}\right]$ eigenvalue
triplets  and $(N+1)\mod3$ single eigenvalues. In section
\ref{sec_pert}, this unfolding behavior will be analytically
explained with the help of perturbational techniques.

\begin{figure}[htb]
 \centering
 \includegraphics[width=6.4cm]{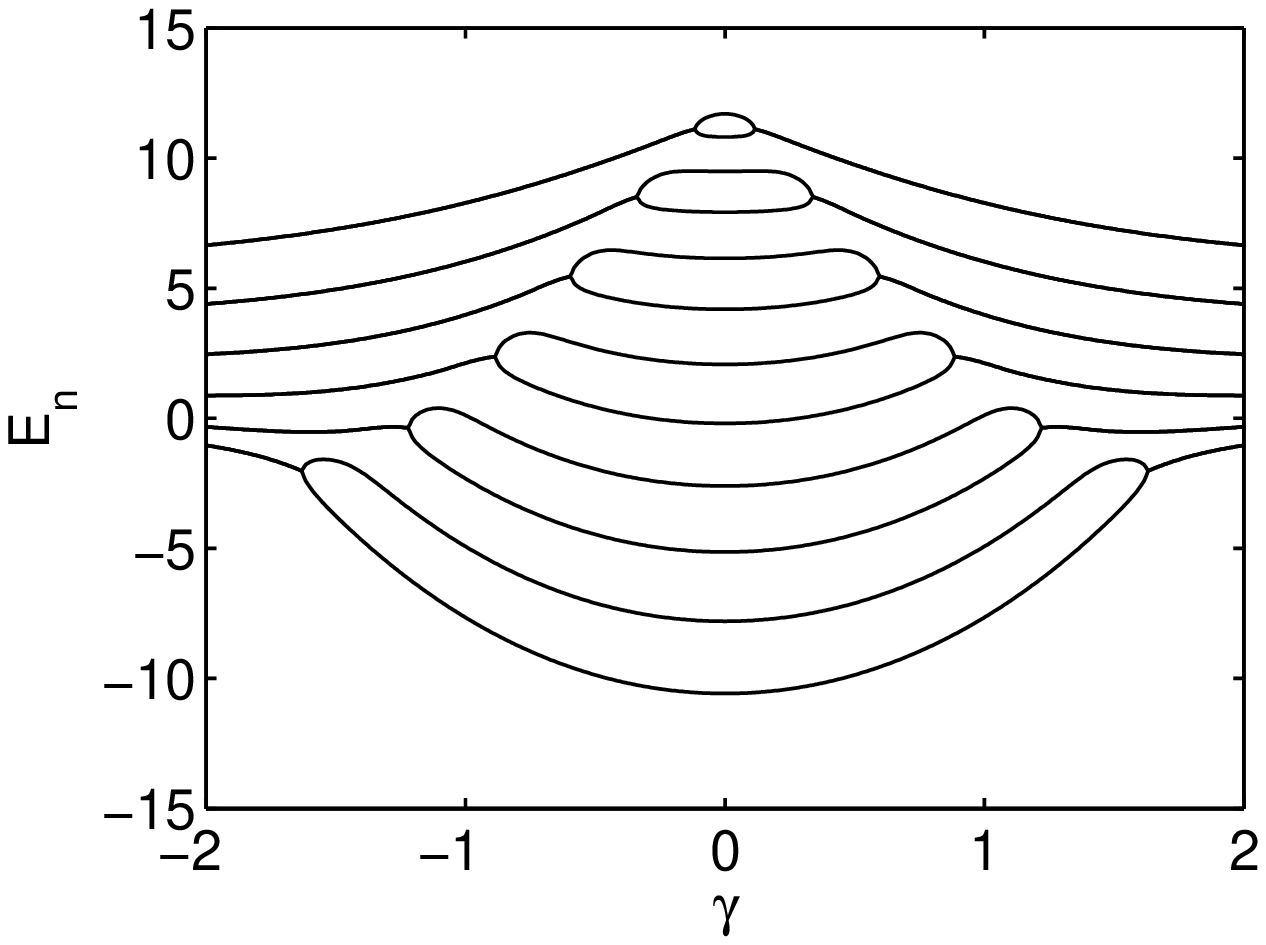}
 \includegraphics[width=6.4cm]{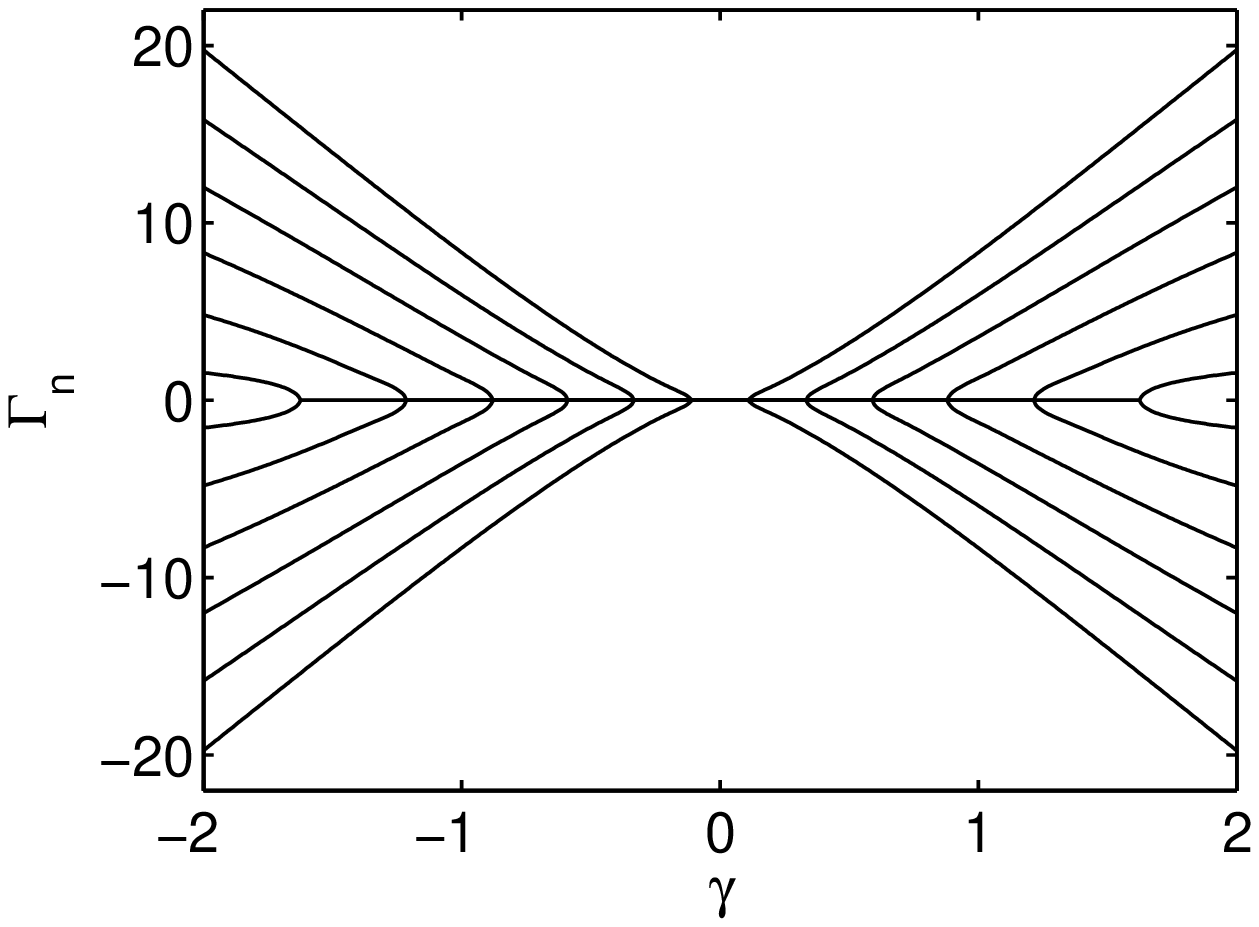}\\
 \includegraphics[width=6.4cm]{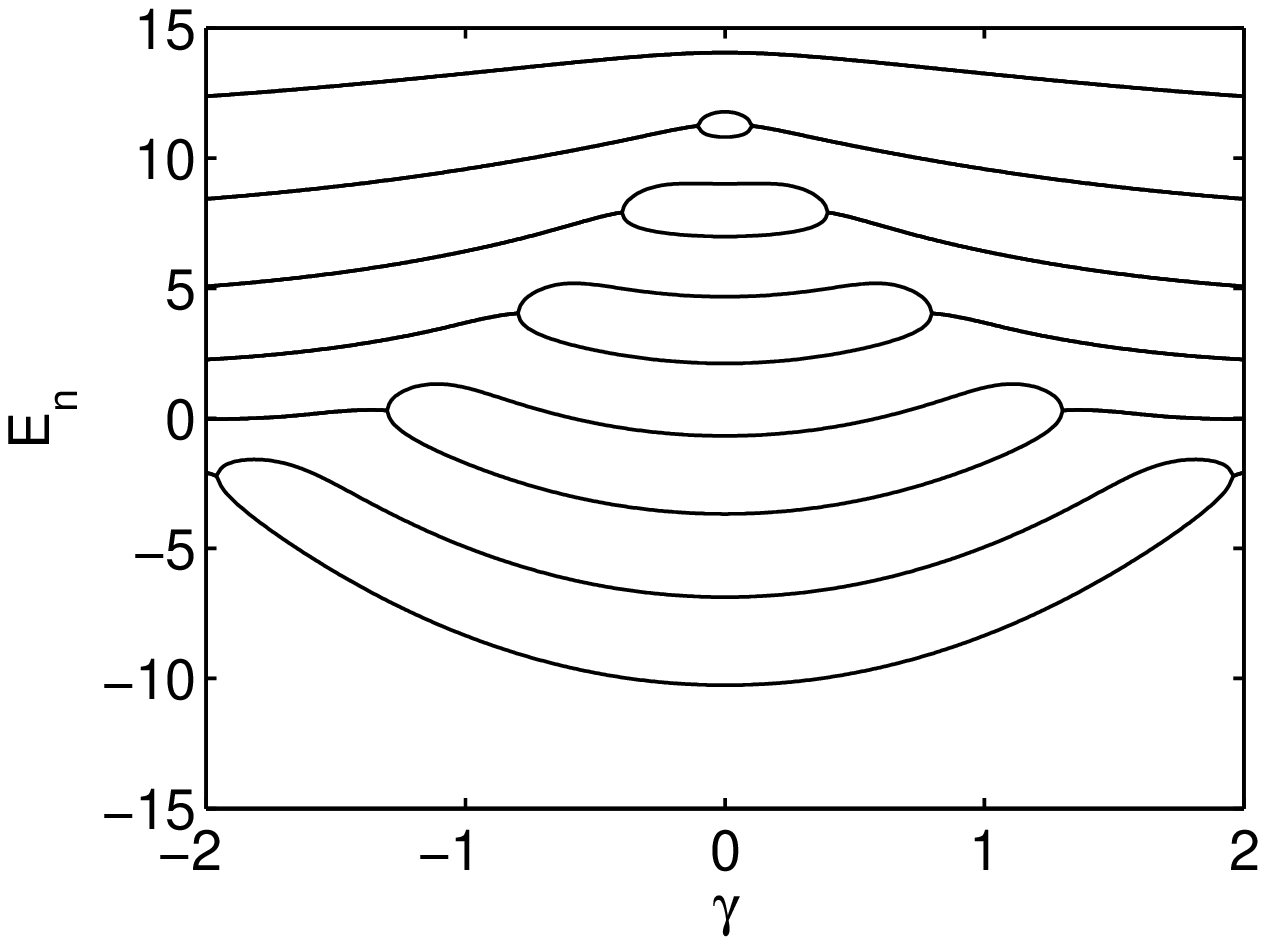}
 \includegraphics[width=6.4cm]{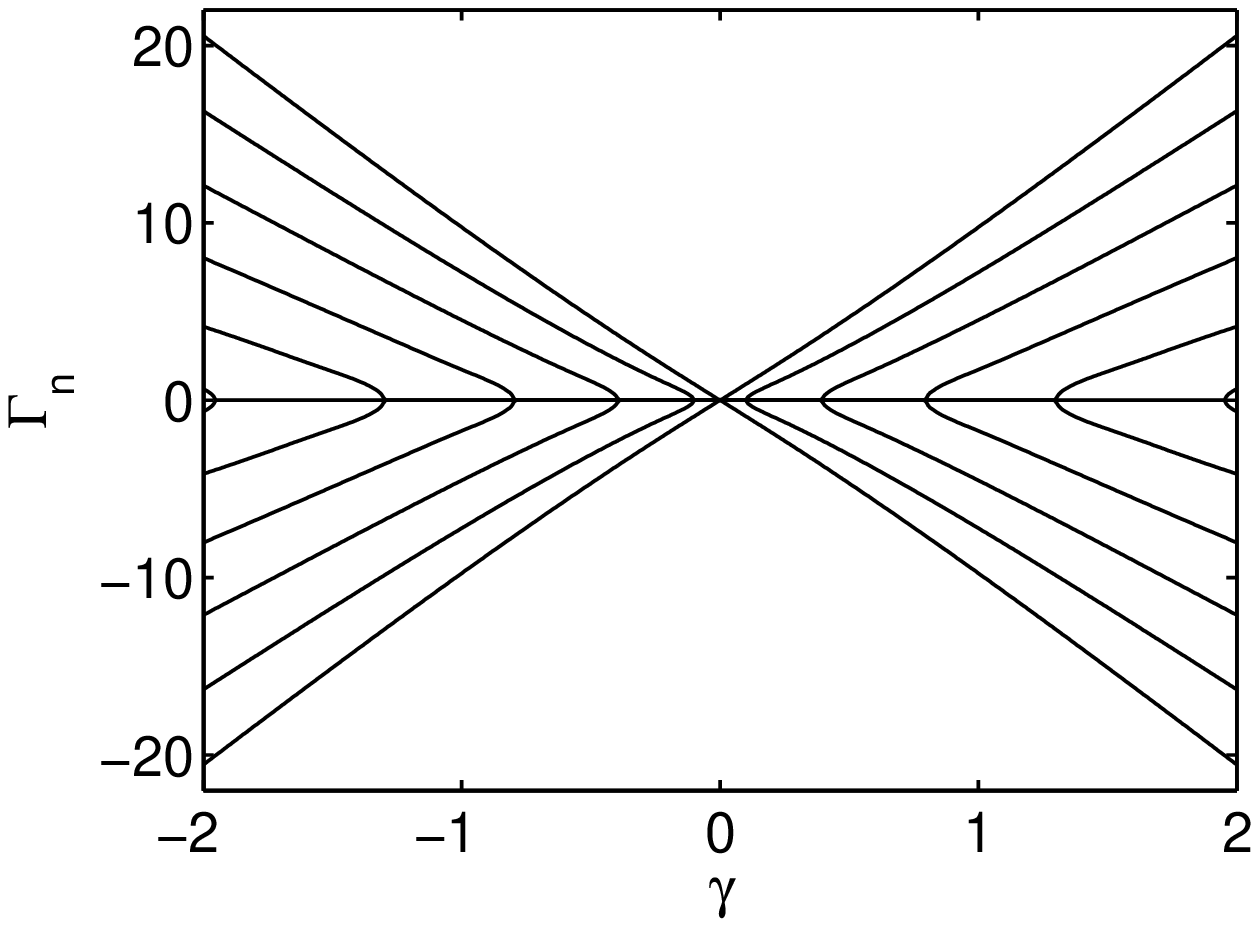}
\caption{\label{fig_EWgamma_c1_2} Real- and imaginary parts of the
 eigenvalues $\lambda_n=E_n-\rmi\Gamma_n$ of the Bose-Hubbard Hamiltonian \rf{pt-ham} as a function of the non-Hermiticity
 $\gamma$ for $v=1$, $N=11$ particles and $c=1/N$ (upper figures) resp. $c=2/N$ (lower figures).}
 \end{figure}
\begin{figure}[!htb]
\begin{center}
\includegraphics[width=6.4cm]{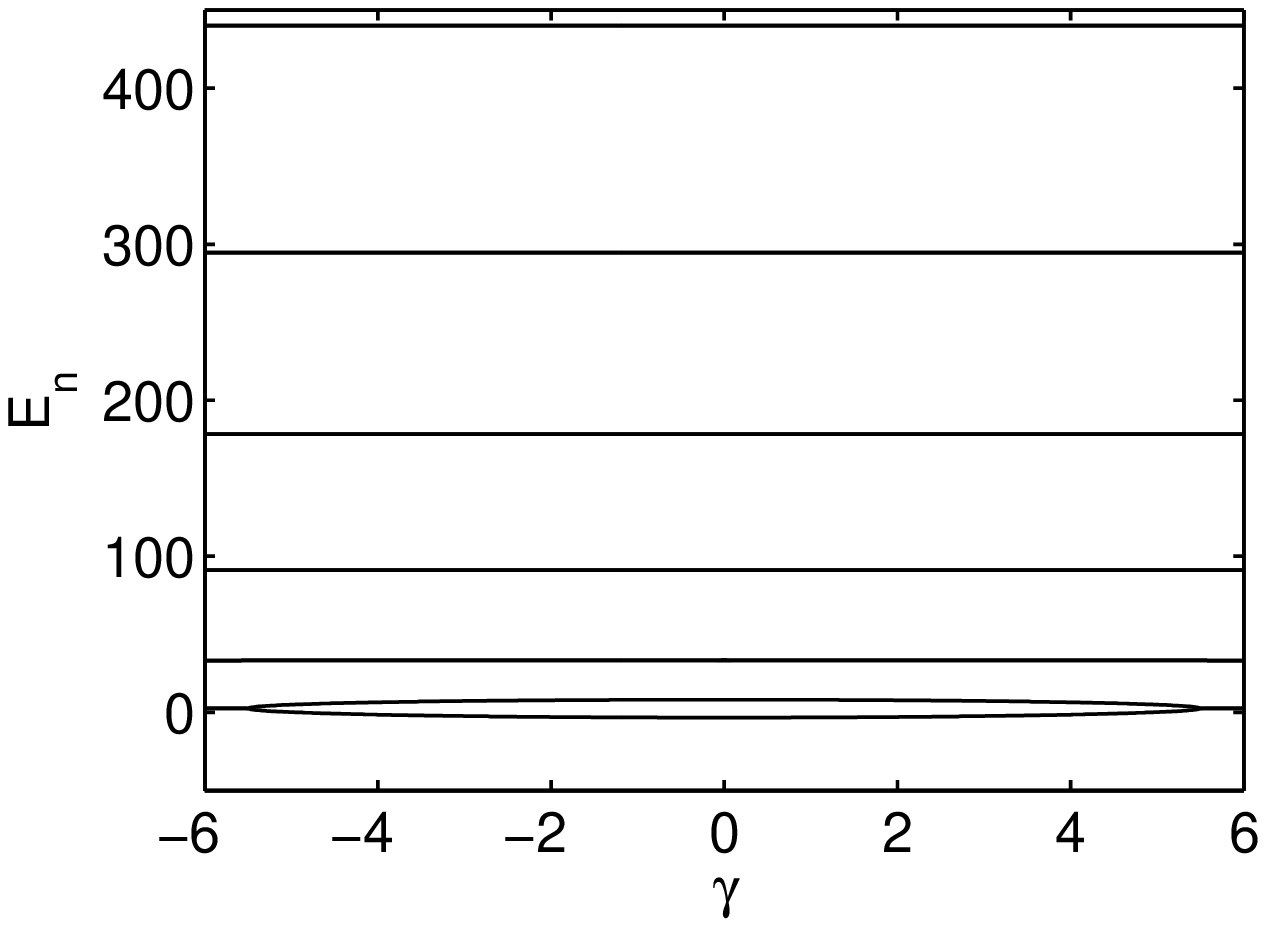}
\includegraphics[width=6.4cm]{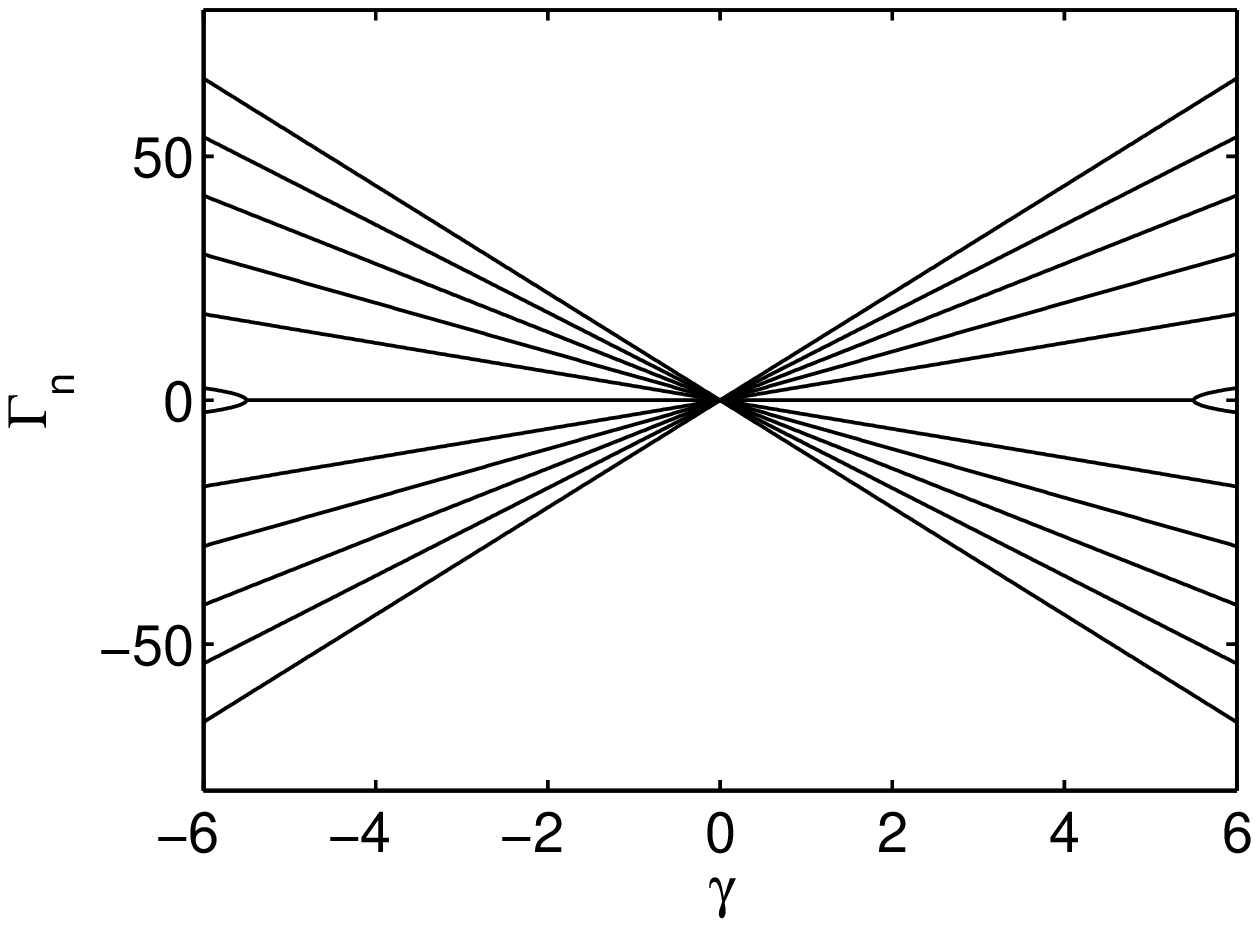}
\caption{\label{Spektrum5} Real- and imaginary parts of the
 eigenvalues $\lambda_n=E_n-\rmi\Gamma_n$ of the Bose-Hubbard Hamiltonian \rf{pt-ham} as a function of the non-Hermiticity
 $\gamma$ for $v=1$, $N=11$ particles and $c=80/N$.}
\end{center}
\end{figure}
\begin{figure}[!htb]
\begin{center}
\includegraphics[width=6.4cm]{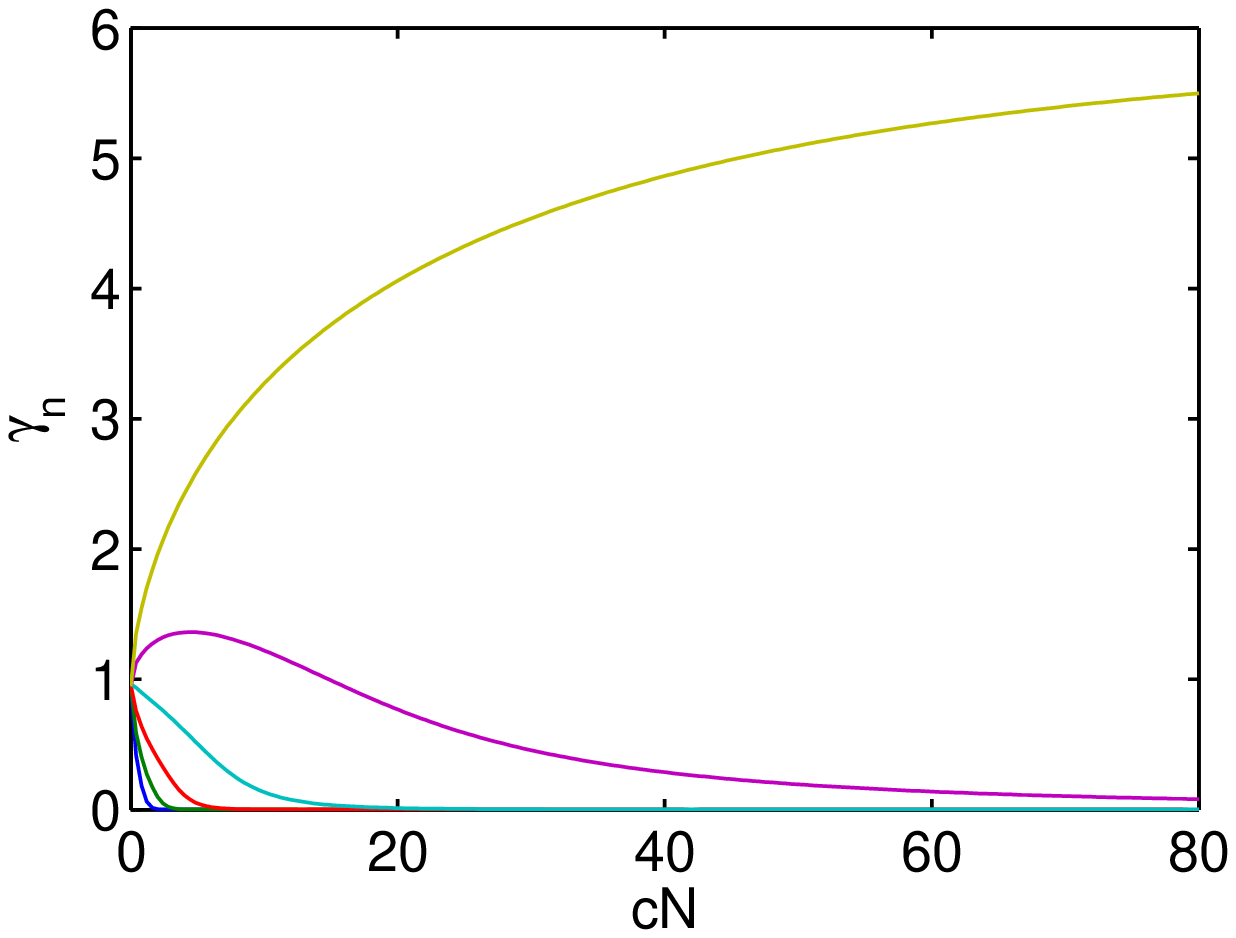}
\includegraphics[width=6.4cm]{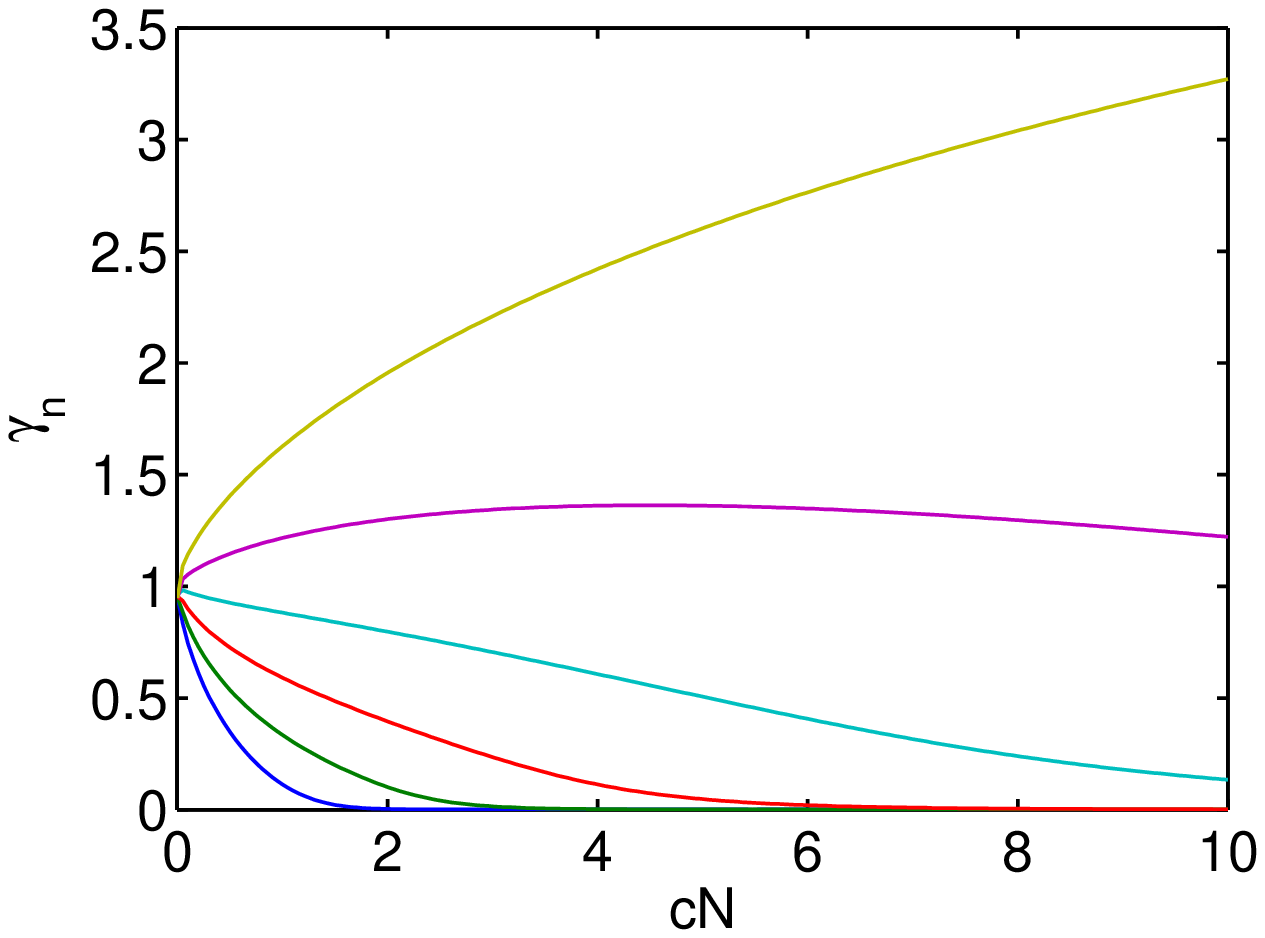}
\caption{\label{EP-positions}(Color online) EP positions in the
$(c,\g)-$half-plane $\g\ge 0$ for fixed $v=1$ and $N=11$. The figure
on the right is a magnification of the small $cN$ region. The EP
positions are numerically approximated as the smallest positive
values of $\gamma$ at which two formally degenerate resonance widths
$\Gamma_{j,k}$ differ by more than $10^{-4}$.}
\end{center}
\end{figure}

After getting a rough qualitative understanding of the unfolding of
the mother EPs at $|\g|=|v|$, $c=0$ we turn now to the system
behavior for intermediate and large interaction strengths $c$.
Figures \ref{fig_EWgamma_c1_2} and \ref{Spektrum5} give an
impression about the corresponding spectral behavior and figure
\ref{EP-positions} about the associated location of the EPs in the
$(c,\g)-$half-plane. Obviously, the movement of the second-order EPs
remains qualitatively the same for intermediate values of $c$: two
of the six EPs connected with the mother EP at $\g=v=1$ move away
from $\g=1$ to positions $\tilde \g_k>1$, whereas the other four
$\tilde \g_k$ remain located within the interval $(0,1)\ni \tilde
\g_k$ and move to smaller $\g$. In the limit of $c\to \infty$ not
only these four EPs tend asymptotically to the limiting value $\g=0$
(see figures \ref{Spektrum5} and \ref{EP-positions}), but also the EP
$\tilde \g_5$ originally located at $\g>1$. The value $\g=0$ itself
is only reached at $c=\infty$, because for $c<\infty$ and $\g=0$ the
Hamiltonian $H$ in \rf{eqn-ham-matrix} is a tridiagonal symmetric
purely real matrix with non-vanishing elements on the superdiagonal.
According to \cite{Wilk} (chapter 5, sections 36 and 37) such
matrices have distinct (nondegenerate) eigenvalues.

Figure \ref{fig_EWgamma_c1_2} clearly shows the shrinking of real
spectral ``bubbles'' which are defined by the intervals $[-\tilde
\g_k,\tilde \g_k]$, $k\le 5$ for increasing $c$. The same fact is
implicitly reflected in the EP behavior in the $(c,\g)-$half-plane
(figure \ref{EP-positions}). The $\g-$positions of the EPs and with
them the sizes of the real spectral ``bubbles'' tend only
asymptotically to zero. By zooming into figures
\ref{fig_EWgamma_c1_2} and \ref{EP-positions} one would observe
still remaining real-energy regions for all spectral branches shown
in figure~\ref{fig_EWgamma_c1_2}. We note that the smallest of those
regions $[-\tilde \g_1,\tilde \g_1]$ defines the sector of exact
$\cP\cT-$symmetry (all eigenvalues are purely real). For
$c\to\infty$ this region together with the other intervals $[-\tilde
\g_k,\tilde \g_k]$, $k\le 5$ shrinks to zero width  (see figure
\ref{Spektrum5}). The shrinkage effect for purely real branches and
the widening of branches with complex conjugate eigenvalues is
similar to the well known effect of eigenvalue complexification for
strong-coupling regimes in Sturm-Liouville type $\cP\cT-$symmetric
models \cite{LT-1,GSZ-squire}.

In contrast to the EP accumulation $\tilde \g_k\to 0$, $k\le 5$, for
$c\to \infty$, the single EP pair $\pm \tilde \g_6$ tends
asymptotically to the limiting values  $\pm \tilde
\g^{(\infty)}_6=\pm 6$. This is the manifestation of a generic
result which holds for systems with odd particle numbers $N$ and
which is derived by perturbation techniques in section \ref{strong
coupling}. It states that for odd $N$ and $c\to \infty$ the lowest
two levels coalesce at EPs located at $\tilde
\g_{\frac{N+1}2}^{(\infty)}=\pm v\frac{N+1}2$. In case of even $N$
all EP positions tend asymptotically to $\g=0$.

Let us restrict our attention (for symmetry reasons
$\g\rightleftharpoons -\g$) to the region $\g\ge 0$. From the
limiting locations of the EPs for $c\to 0$ and $c\to \infty$ and the
numerical results for $c\in [0,\infty)$ we conclude that the second
order EPs are located on two-dimensional hypersurfaces
$\cV_n=\{(\g,v,c)\in \cM: \ \g=\g_n(v,c)\}$ in the three-dimensional
parameter space $ \cM\ni (\g,v,c)$. For $c=0$ these surfaces $\cV_n$
coalesce at a common  line $\bigcap_n \cV_n=\{ (\g,v,c)\in\cM: \ \g=
v,c=0\}$. In the opposite limit of $c\to\infty$ all but one of the
surfaces (in case of $N$ odd) tend asymptotically to the plane
$\cN_0=\{(\g,v,c)\in \cM: \ \g=0\}$. The remaining  surface $\cV_{
\frac{N+1}2}$ asymptotically approaches the plane
$\cN_+=\{(\g,v,c)\in \cM: \ \g= v\frac{N+1}2\}$. Hence, the
stratified manifold $\cV\subset \cM$ of EP degenerations for the
present model comprises the two-dimensional surfaces $\cV_n$ which
intersect (coalesce) at the line $\bigcap_n \cV_n$
\be{c0-12a}
\cV=\bigcup_n \cV_n
\ee
and its mirror images under the symmetry transformation $\g
\rightleftharpoons -\g$.

Summarizing the numerical findings we conclude that although the
influence of the particle interaction might stabilize some of the
energy levels, i.e. it might shift the occurrence of their imaginary
parts to higher values of the non-Hermiticity parameter $\gamma$,
the position of the first EPs  monotonically decreases with
increasing interaction strength $c$. Therefore the interaction
always shrinks the region of unbroken PT symmetry. In the limit of
infinitely strong interaction even an arbitrarily small complex
perturbation destroys the reality of the spectrum.

\section{Perturbative results}
\label{sec_pert}

In this section we will investigate some of the features of the
spectrum of the Hamiltonian \rf{pt-ham} for different limiting cases analytically
using perturbational techniques.

At first we will focus on the unfolding of the $(N+1)$th-order
EP due to a weak particle interaction $c$. Since usual
Rayleigh-Schr\"odinger perturbation theory breaks down for
non-Hermitian Hamiltonians in the vicinity of EPs of
the unperturbed operator, we resort to the Puiseux-Newton resp.
Newton-polygon method \cite{baumg,chebotarev,trenogin}, a
perturbative technique for the roots of polynomials which works in
the vicinity of degeneracies as well as around simple values. This
method will allow us to analytically verify the triplet unfolding of
the mother EP we found numerically in section \ref{sec_num}.

Finally, in the second part of this section, we will use standard
Rayleigh-Schr\"odinger perturbation  to understand the behavior of
the spectrum in the strong-interaction regime.

\subsection{The limit of weak interaction\label{weak coupling}}

Here we are going to analyze the unfolding of the $(N+1)$th-order EP
when the interaction is switched on perturbatively, i.e., we
consider the spectral behavior of the Hamiltonian~\rf{pt-ham} at the
EP $\gamma=v$
\be{h-1}
H=2v(L_x-iL_z)+2cL_z^2
\ee
for small interaction $0\le|c|\ll |v|/N$.

As a first step, we $SU(2)-$rotate the Hamiltonian \rf{h-1} with the
help of eq. \rf{rot-x} and $\t=\pi/2$ into the more convenient form
\ba{h-2}
\tilde H&=&2v(L_x-iL_y)+2cL_y^2\nn\\
&=&2vL_--\frac c2(L_+-L_-)^2\nn\\
&=:&2vL_--\frac c2\left(L_+^2 -L_0+L_-^2\right),\qquad
L_0:=L_+L_-+L_-L_+\,.
\ea
In the particle number (angular momentum $l$) representation this
Hamiltonian has a band diagonal structure of the type
\be{h-3}
\tilde H=\left(
    \begin{array}{ccccccccc}
      * & 0 & * & 0 & 0&\cdots &\cdots & 0&0 \\
      * & * & 0 & * & 0& \cdots & \cdots & 0&0 \\
      * & * & * & 0 & *&\cdots & \cdots &0&0 \\
      0 & * & * & * & 0& \cdots & \cdots&0&0 \\
      \vdots & \ddots & \ddots & \ddots & \ddots & \ddots& \ddots&\vdots & \vdots\\
       0& 0 & \cdots & * &  *& * &0& *&0\\
        0 &0  & \cdots & 0 &  *& * &*& 0&*\\
        0 &0  & \cdots & 0 &  0& * &*& *&0\\
        0 &0  & \cdots & 0 &  0& 0 &*& *&*\\
    \end{array}
  \right),
\ee
where the second super- and subdiagonals are generated by the
$L_+^2$ and $L_-^2$ terms and the diagonal by $L_0$. The
perturbation matrix is an upper $3-$Hessenberg matrix, i.e. a matrix
with only zero entries below the three subdiagonals (including the main
diagonal). Therefore the results of \cite{ma} apply, where the
unfolding of the eigenvalues of Jordan blocks $J_n(\lb_0)$ under
perturbations by general upper $k-$Hessenberg matrices has been
analyzed. It has been shown that an $n$th-order EP typically splits
into $p$ rings of size $k$ and one of size $r$ (if $r\neq 0$), where
$n=pk+r$, $p=\left[\frac nk\right]$ and $r=n \mod p$. This means
that for a small coupling parameter $|c|\ll |v|/N$ the EP will unfold
as
\ba{h-3a}
\lb_{q,j}&=&a_q e^{i 2\pi j/k} c^{1/k}+o(c^{1/k}),\qquad
j=0,\ldots,k-1\quad q=1,\ldots, p\nn\\
\lb_{0,l}&=&a_0 e^{i 2\pi l/r} c^{1/r}+o(c^{1/r}),\qquad l=0,\ldots,
r-1.
\ea
The coefficients $a_q$ are specific model dependent constants whose
moduli $|a_q|$ define the scaling of the ring radii. The specific
$n=pk+r$ splitting behavior generalizes the well known results for
the generic case $(p=1,r=0)$ where the degenerate eigenvalue at the
EP splits into a single ring of size $n$, i.e. where (in suitable
reparametrization) $\lb^n=c$ holds, with the ring $\lb=e^{i2\pi
j/n}c^{1/n}$,\quad $j=0,\ldots,n-1$, of size $n$ as obvious
solutions. Applying the results of \cite{ma} to the unfolding of our
EP configuration we expect the formation of
$p=\left[\frac{N+1}3\right]$ rings of size $k=3$ and possibly of a
single ring of size $r=1$  (single eigenvalue) or $r=2$ (eigenvalue
pair) depending on the concrete dimension $N+1$ and $r=(N+1)\mod p$.

Subsequently, we present an explicit derivation of this behavior
which makes use of the specific structure of our model. The analysis
will be based on the characteristic polynomial of the matrix
Hamiltonian $\tilde H$ \rf{h-2} as function of the interaction strength $c$,
\be{h-4}\fl
\chi_{\tilde H}(\lb,c)=\det\left(\lb I-\tilde
H\right)=-\sum_{k=0}^Mp_{M-k}(c)\lambda^k, \qquad M:=N+1
\ee
which we will study with the help of the Newton-polygon technique
\cite{trenogin,baumg}. This will allow us to derive the dominant
fractional power $\mu$ of the unfolding $\lb=a_0 c^\mu+o(c^\mu)$ of
the $(N+1)$th-order EP.

We start our analysis by noticing that the characteristic polynomial
of any $\cP\cT-$symmetric matrix Hamiltonian $H$ has purely real
coefficients. This  follows straightforwardly  from the fact that
any $\cP\cT-$symmetric operator can be represented as a purely real
(possibly infinite dimensional) matrix \cite{bender-berry}.

The  coefficients  $p_{M-k}(c)$ in \rf{h-4} can be recursively
obtained with the help of the Le Verrier-Faddeev method
\cite{gantmacher} as
\ba{h-5} p_k=-\frac
1k\sum_{j=1}^k s_jp_{k-j}, \quad k=1,\ldots, M,\\
 s_k:=\Tr(\tilde H^k),
\qquad p_0=-1\,.\nn
\ea
For our purpose it is sufficient to extract the structure of the
coefficients $p_k(c)$ as polynomials in $c$. As basic input we first
derive the corresponding traces $s_k(c)=\Tr(\tilde H^k)$. These
traces act in the angular momentum representation \rf{ang-mom-4} or,
equivalently,  in the monomial representation \rf{ang-mom-1}. This
means that only terms in $\tilde H^k$ contribute to $\Tr(\tilde
H^k)$ which leave the angular momentum mode number $m$, and with it the monomial
power, unchanged. Hence it will hold , e.g., $\Tr(L_\pm^k)=0$
$\forall k\in \ZZ^+$, i.e. $k> 0$, as well as $\Tr(L_+^kL_-^j)=0$
$\forall k\neq j \in \ZZ^+$, but $\Tr(L_+^kL_-^k)\neq 0$. In order
to extract the power structure of $s_k(c)$ in $c$ we may relate to
$L_\pm$ their auxiliary commutative symbols $L_+\approx \xi$,
$L_-\approx \xi^{-1}$. The terms $\tilde H^k$ can then be associated
with the multinomials
\ba{h-6} \tilde H^k&\approx &
\left[v\xi^{-1}+c(\xi-\xi^{-1})^2\right]^k\nn\\
&\approx & \sum_{l=0}^k\left({k \atop
l}\right)v^{k-l}\xi^{l-k}c^l\sum_{i=0}^{2l}\left({2l \atop
i}\right)(-1)^i \xi^{2(i-l)}
\ea
(we omitted the irrelevant pre-factors $2$ and $-1/2$ in front of
$v$ and $c$) where only the constant $\xi^0-$terms will contribute
to the trace $\Tr(\tilde H^k)$. With the notation $j:=i-l$ we find
the $\xi^0-$terms as $\xi^0=\xi^{l-k+2j}$. Combining the
corresponding constraint $k-l=2j$ with the inequalities $k\ge l\ge
0$, $2l\ge i\ge 0$ one obtains $j\le l=k-2j$ and, hence, the
condition $k\ge 3j$ or $j\le\left[\frac k3\right]$.

As result, we find the structure of the traces as
\be{h-7}
s_k(c,v)=\Tr(\tilde H^k)=\sum_{j=0}^{\left[\frac
k3\right]}a_j^{(k)}c^{k-2j}v^{2j}.
\ee
The constant coefficients $a_j^{(k)}$ depend only on $k$ and the
particle number $N$ (resp. the angular momentum $l$) and can be calculated
with the help of general trace formulae for polynomials of angular
momentum operators as discussed, e.g., in \cite{subra-trace}. In our
subsequent qualitative analysis we are only interested in the
behavior of $s_k(c,v)$ as polynomial in $c$ so that the concrete
values of the non-vanishing coefficients $a_j^{(k)}$ are irrelevant.

The trace formula \rf{h-7} allows us to prove the following

\begin{Th}
The coefficients $p_k$ have the structure
\be{h-8}
p_k=\sum_{j=0}^{[\frac{k}{3}]}d_j^{(k)}c^{k-2j}v^{2j}
\ee
where $d_j^{(k)}$ are real constants which in general do not vanish.
\end{Th}

\begin{proof}
The theorem is true for $k=0$ and $k=1$ where \rf{h-5} and \rf{h-7}
imply
\be{h-8a}
p_0=-1,\qquad p_1=-p_0s_1=s_1=a_0c^1\,.
\ee
{}From \rf{h-5} and \rf{h-7} one finds \rf{h-8} by induction:
\ba{h-9}
p_{n+1}&=&-\frac 1{n+1}\sum_{k=1}^{n+1} s_k p_{n+1-k}\nn\\
&=&-\frac 1{n+1}\sum_{k=1}^{n+1}\sum_{l=0}^{\left[\frac
k3\right]}\sum_{j=0}^{[\frac{n+1-k}{3}]}a_l^{(k)}d_j^{(n+1-k)}c^{n+1-2(l+j)}v^{2(l+j)}\label{h-9a}\\
&=&\sum_{r=0}^{[\frac{n+1}{3}]}b_r^{(n+1)}c^{n+1-2r}v^{2r}\label{h-9b}\\
b_r^{(n+1)}&:=&-\frac 1{n+1}\sum_{k=1}^{n+1}\sum_{l=0}^{\left[\frac
k3\right]}\sum_{j=0}^{[\frac{n+1-k}{3}]}a_l^{(k)}d_j^{(n+1-k)}\delta_{r,l+j}\,.\nn
\ea
In passing from \rf{h-9a} to \rf{h-9b} we used the fact that the
total summation in \rf{h-9a} goes over the three-dimensional
discrete volume $\Omega=\{k,l,j:\ 1\le k\le n+1,\ 0\le l
\le\left[\frac k3\right],\ 0\le j \le [\frac{n+1-k}{3}]\}$ and that
$\Omega$ is recovered by slicing it along fixed $r=l+j$ and summing
over the two-dimensional slices.
\end{proof}

Summarizing the above results we conclude that for a given index $k$
it is sufficient to use the single index $j$ as basic counting index,
whereas the power of $c$ in the polynomial terms is defined by the
derived value $l$ according to the relation:
\ba{h10b}
l&=& k-2j\qquad j\le\left[\frac{k}{3}\right].
\ea
As an illustration we list the $c-$dependence of the first six
non-trivial coefficients $p_k$, $k=1,\ldots, 6$ of the
characteristic polynomial $\chi_{\tilde H}(\lb,c)$. According to
\rf{h-8} these coefficients contain terms of the following
$c^lv^{2j}-$monomial types
\ba{h11}
 k=1&\qquad &j=0,\quad l=1,\qquad c^1v^0\nn\\
 k=2&\qquad &j=0,\quad l=2,\qquad c^2v^0\nn\\
 k=3&&j=0,\quad l=3,\qquad c^3v^0\nn\\
 &&j=1,\quad l=1,\qquad c^1v^2\nn\\
 k=4&&j=0,\quad l=4,\qquad c^4v^0\nn\\
 &&j=1,\quad l=2,\qquad c^2v^2\nn\\
 k=5&&j=0,\quad l=5,\qquad c^5v^0\nn\\
 &&j=1,\quad l=3,\qquad c^3v^2\nn\\
 k=6&&j=0,\quad l=6,\qquad c^6v^0\nn\\
 &&j=1,\quad l=4,\qquad c^4v^2\nn\\
 &&j=2,\quad l=2,\qquad c^2v^4\,.
\ea

With the concrete monomial terms \rf{h11} at hand, it is now an easy
task to obtain the leading (dominating) exponent $\mu_1$ in the
power series expansion
\be{h12}
\lb(c)=e_1 c^{\mu_1}+e_2 c^{\mu_2}+o(c^{\mu_2}), \quad
0<\mu_1<\mu_2<\cdots, \quad e_i\neq 0
\ee
for small $|c|\ll |v|/N$. The method that we use is known as
Puiseux-Newton diagram technique or Newton-polygon technique
\cite{baumg,chebotarev,trenogin} and can be summarized as follows.
One starts by substituting the ansatz \rf{h12} with still unknown
exponents $\mu_i$ and coefficients $e_i$ into the characteristic
polynomial \rf{h-4}
\ba{h13}\fl
\chi_{\tilde H}[\lb(c),c]&=&\lb^M-p_1(c)\lb^{M-1}-\cdots
-p_{M-1}(c)\lb -p_M(c)=0\nn\\
&=&\left(e_1
c^{\mu_1}+\cdots\right)^M-\left(f_{M-1}c^{a_{M-1}}+\cdots\right)\left(e_1
c^{\mu_1}+\cdots\right)^{M-1}-\nn\\
&&-\cdots-\left(f_{1}c^{a_{1}}+\cdots\right)\left(e_1
c^{\mu_1}+\cdots\right) - \left(f_0 c^{a_0}+\cdots\right)=0.
\ea
The dots in the various terms denote contribution of higher powers
in $c$. For later convenience the lowest $c-$powers in the
coefficients $p_k$ are numbered in reversed order as
\be{h13a}
p_k(c)=f_{M-k}c^{a_{M-k}}+o(c^{a_{M-k}})\,.
\ee
Since $\chi_{\tilde H}[\lb(c),c]$ as polynomial in $c$ has to
vanish, $\chi_{\tilde H}[\lb(c),c]=0$, it should contain at least
two terms in each power in $c$ so that these terms can compensate
one another. Taking into account that terms in the lowest power in
$c$ are the most dominating ones, one has to search for these
lowest-power-terms and to fix the still undefined $\mu_i$ in such a
way that these terms compensate. After fixing the minimal $\mu_1$
one repeats the process for the next greater $\mu_2$ and so on. In
this way one can iteratively obtain the series expansion \rf{h12} up
to any required precision. The validity of the perturbative results
is of course limited by the finite convergence radius of the
perturbation series \rf{h12}.

\begin{figure}[htb]
\begin{center}
 \includegraphics[scale=0.35]{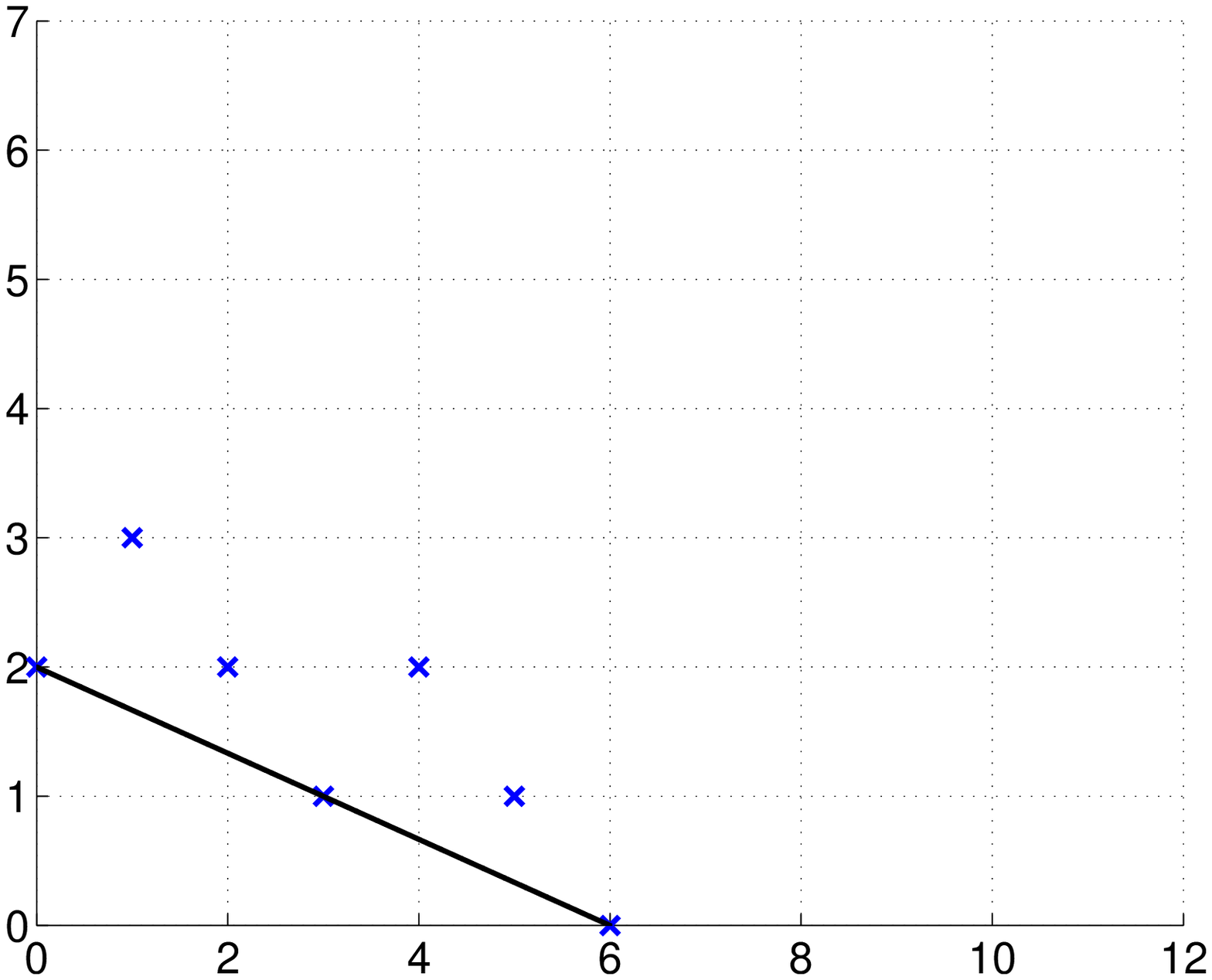}
 \includegraphics[scale=0.35]{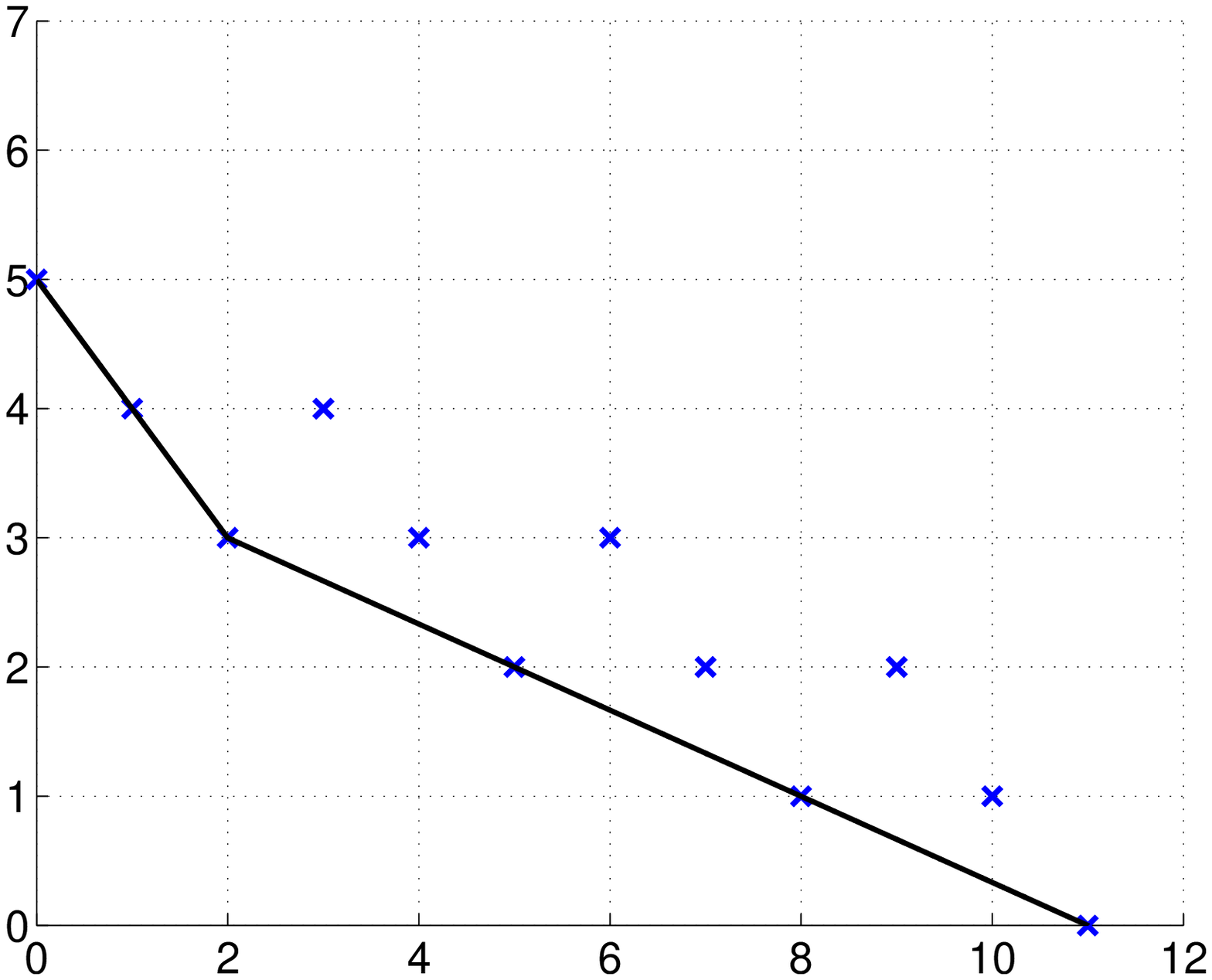}
 \caption{\label{Puiseux}Puiseux-Newton-diagram for $N=5$ particles (left) and $N=10$ particles (right).}
\end{center}
\end{figure}

Returning to \rf{h13}, one sees that $\mu_1$ as minimal exponent
should be defined from the monomials
\be{h14}\fl
e_1^M c^{M\mu_1}, \ f_{M-1}e_1^{M-1}c^{a_{M-1}+(M-1)\mu_1}, \
f_{M-2}e_1^{M-2}c^{a_{M-2}+(M-2)\mu_1},\ldots, f_0 c^{a_0}
\ee
by pairwise identifying their powers
\be{h15}
a_k+ k\mu_{j,k}=a_j+ j\mu_{j,k}, \qquad j\neq k, \quad j,k=0,\ldots,
M.
\ee
In order to single out the relevant values of $\mu_1$ one associates
to each power $a_k+k\mu_{j,k}$ a point $A_k=(k,a_k)$ in the
$(k,a_k)-$plane so that
\be{h15a}
\mu_{j,k}=-\frac{a_k-a_j}{k-j}
\ee
is just the sign-inverted slope of the line connecting the points
$A_j$ and $A_k$. As shown, e.g., in
\cite{baumg,chebotarev,trenogin}, the possible values of $\mu_1$ can
then be identified with those $\mu_{j,k}$ whose lines form the lower
boundary of the convex hull of the points $A_k$. The corresponding
graph is the so-called Newton polygon. Examples are depicted in
figure~\ref{Puiseux}.
Points above this lower boundary of the convex hull will only
contribute to higher order approximations.

The coefficients $e_1$ in the series expansion \rf{h12} can be
obtained from reduced polynomials which are built from those
leading-order terms of the characteristic polynomial \rf{h13} which
correspond to points $A_k$ located on the same lines of the
Newton-polygon. In general, the lines comprise more than two points
as it is visible, e.g., in figure \ref{Puiseux}. Here for $N=5$ the
$(\mu_1=1/3)-$line comprises three points and for $N=10$ the
$(\mu_1=1)-$line also comprises three points and the
$(\mu_1=1/3)-$line four points.

Applying the described Puiseux-Newton technique to our concrete
characteristic polynomial we read off from eq. \rf{h-8} that the
lowest $c-$powers in the coefficients $p_k(c)$ have the form
$d^{(k)}_{\left[\frac k3\right]}c^{k-2\left[\frac k3\right]}$. With
$M=N+1$ and the reverse numbering \rf{h13a} we find the
relevant points $A_k$ in the $(k,a_k)-$plane as
\be{h16}\fl
A_{N+1-k}=\left(N+1-k,k-2\left[\frac{k}{3} \right]\right)\,,\qquad
k=0,\ldots, N+1.
\ee
In case of $N=5$ this gives, for example, the seven points
\ba{h17}
&&A_0=(0,2), \quad A_1=(1,3), \quad A_2=(2,2),\quad A_3=(3,1)\nn\\
&&A_4=(4,2), \quad A_5=(5,1), \quad A_6=(6,0)
\ea
which are depicted in the left graphics of figure~\ref{Puiseux}.
Both of the two graphics in this figure show a typical
modulo-three-ratchet-structure of the points $A_k$. The lower
boundary of the convex hull of these points is always formed by one
(long) line of slope $-1/3$ (and corresponding $\mu_1=1/3$) which
connects $\left[\frac{N+1}{3}\right]+1$ points $A_k$ and possibly a
second (short) line of slope $-1$ (and $\mu_1=1$) which connects the
first two or three leftmost points. We arrive at the result that in
our model only dominant eigenvalue scalings of the type
\be{h19}
\lb\propto c^{1/3},\qquad \lb\propto c
\ee
are possible. Obviously, these eigenvalues will form rings of size
three (triplets) and of size one (single eigenvalues). The specific
size-one rings can be considered as atypical cases in the scheme of
\cite{ma}. They can be attributed to the specific substructure of
the Hessenberg type perturbation matrix. Moreover these $\lb\sim c$
terms can be identified as higher-order corrections --- what is
clearly visible by setting $c^{1/3}=:\e$ so that $c=\e^3$. This is a
specific output of the Newton-diagram technique which yields the
dominant terms for all roots of a polynomial. In our concrete case
this means that the $3-$rings correspond to dominant lowest-order
scaling $c^{1/3}$, whereas the single eigenvalues remain unperturbed
in this lowest order. Their dominant terms start only with the
higher-order $\lb\sim c$ contributions.

Let us now illustrate the general theoretical results by more
explicit calculations for models with $N=5$ and $N=10$ particles.
For the $N=5$ model the corresponding Hamiltonian has the form
\begin{equation}\label{example1}\fl
H= \left(
\begin{array}{cccccc}
 -5 {\rm i} \gamma+ \frac{25}{2} c & \sqrt{5} v & 0 & 0 & 0 & 0\\
 \sqrt{5} v & -3{\rm i}\gamma+\frac{9}{2}c & 2\sqrt{2} v & 0 & 0 & 0\\
 0 & 2\sqrt{2}v & -{\rm i}\gamma+\frac{1}{2}c & 3v & 0 & 0\\
 0 & 0 & 3v & {\rm i}\gamma+\frac{1}{2}c & 2\sqrt{2}v & 0\\
 0 & 0 & 0 & 2\sqrt{2}v & 3{\rm i}\gamma+\frac{9}{2}c & \sqrt{5}v\\
 0 & 0 & 0 & 0 & \sqrt{5}v & 5{\rm i}\gamma+\frac{25}{2}c
\end{array}
\right).
\end{equation}
In case of $\g=v$ the characteristic polynomial reads
\begin{eqnarray}\fl
0&=&\det(H-\lambda I)\nonumber \\
\fl&=&\lambda^6-35c\lambda^5+\frac{1743}{4}c^2 \lambda^4+\left(-\frac{4645}{2}c^3+448v^2c\right)\lambda^3\nonumber\\
\fl&&\:+\left(\frac{82831}{16}c^4-6112v^2c^2\right)\lambda^2+\left(-\frac{58275}{16}c^5+27280v^2c^3\right)\lambda\nonumber\\
\fl&&\:+\frac{50625}{64}c^6-30600v^2c^4+6400v^4c^2.
\end{eqnarray}
Obviously, $\lb=0$ is the only root for $c=0$. For $|c|> 0$
the coordinates of the points \rf{h17} in the Puiseux-Newton diagram
are given by the exponent of $\lambda$ and the minimal exponent of
$c$ of each summand. Under the conditions described above one finds
one straight line with the slope $-1/3$ and associated $\mu_1=1/3$.
The three summands corresponding to the three points on this
straight line must compensate each other
\begin{equation}
 6400 c^2 v^4+448v^2c \lambda^3+\lambda^6=0,\quad
 \lambda=e_1c^{1/3}+\ldots\,.
\end{equation}
This yields two solution triplets for the coefficient $e_1$:
\begin{eqnarray}\label{coeff_ex}\fl
&&e_1^{(1)}=-\sqrt[3]{14.77} v^{\frac{2}{3}},\phantom{e^{-{\rm i} \frac{\pi}{3}}}
\qquad e_1^{(4)}=-\sqrt[3]{433.23} v^{\frac{2}{3}},\nonumber \\
\fl&&e_1^{(2)}=\phantom{-}\sqrt[3]{14.77} v^{\frac{2}{3}} e^{{\rm i} \frac{\pi}{3}},
\phantom{-}\qquad e_1^{(5)}=\phantom{-}\sqrt[3]{433.23} v^{\frac{2}{3}} e^{{\rm i} \frac{\pi}{3}},\nonumber \\
\fl &&e_1^{(3)}=\phantom{-}\sqrt[3]{14.77} v^{\frac{2}{3}} e^{-{\rm
i} \frac{\pi}{3}},\qquad  e_1^{(6)}=\phantom{-}\sqrt[3]{433.23}
v^{\frac{2}{3}} e^{-{\rm i} \frac{\pi}{3}}.
\end{eqnarray}

\begin{figure}[!htb]
\begin{center}
 \includegraphics[width=6.4cm]{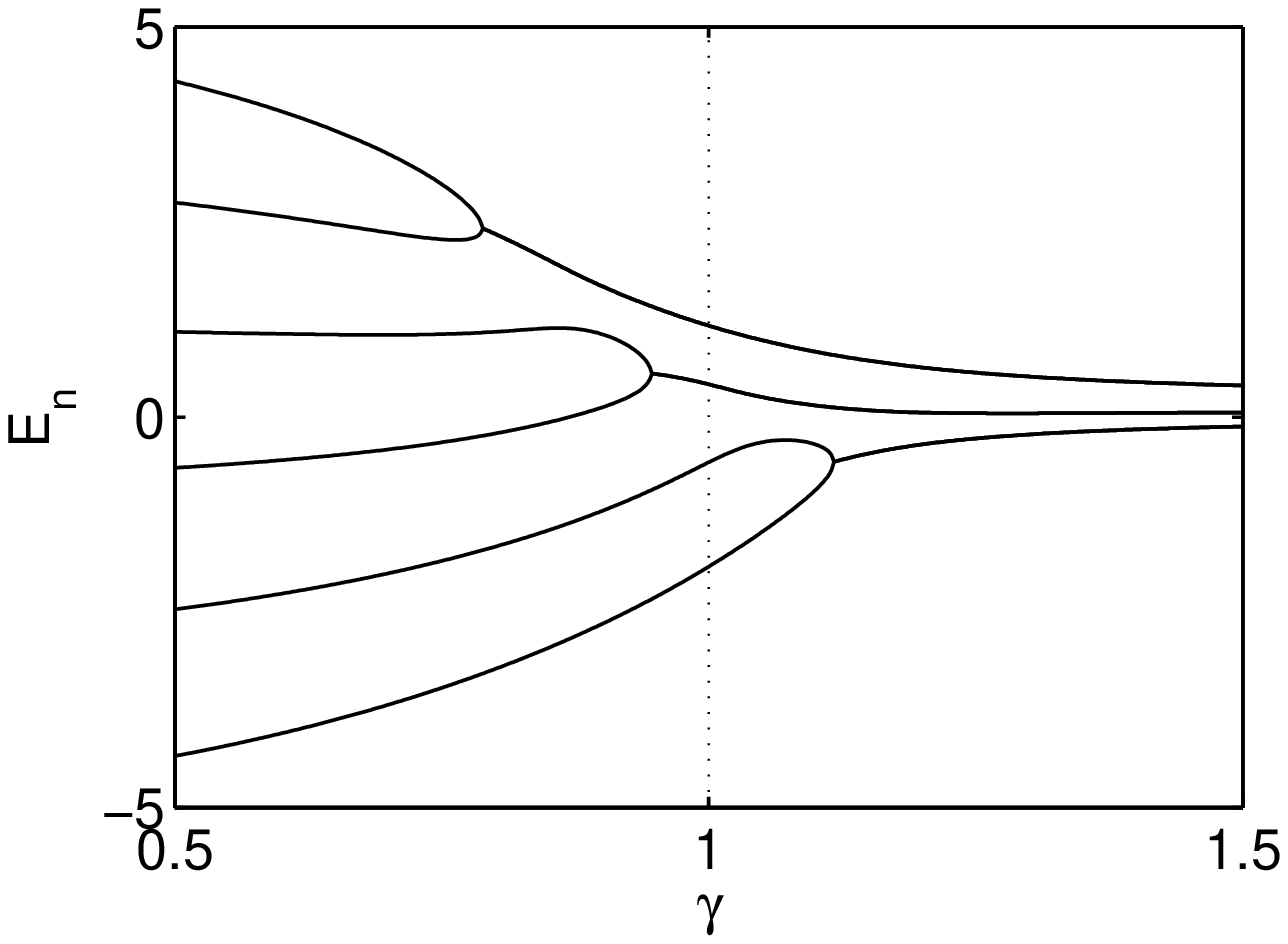}
 \includegraphics[width=6.4cm]{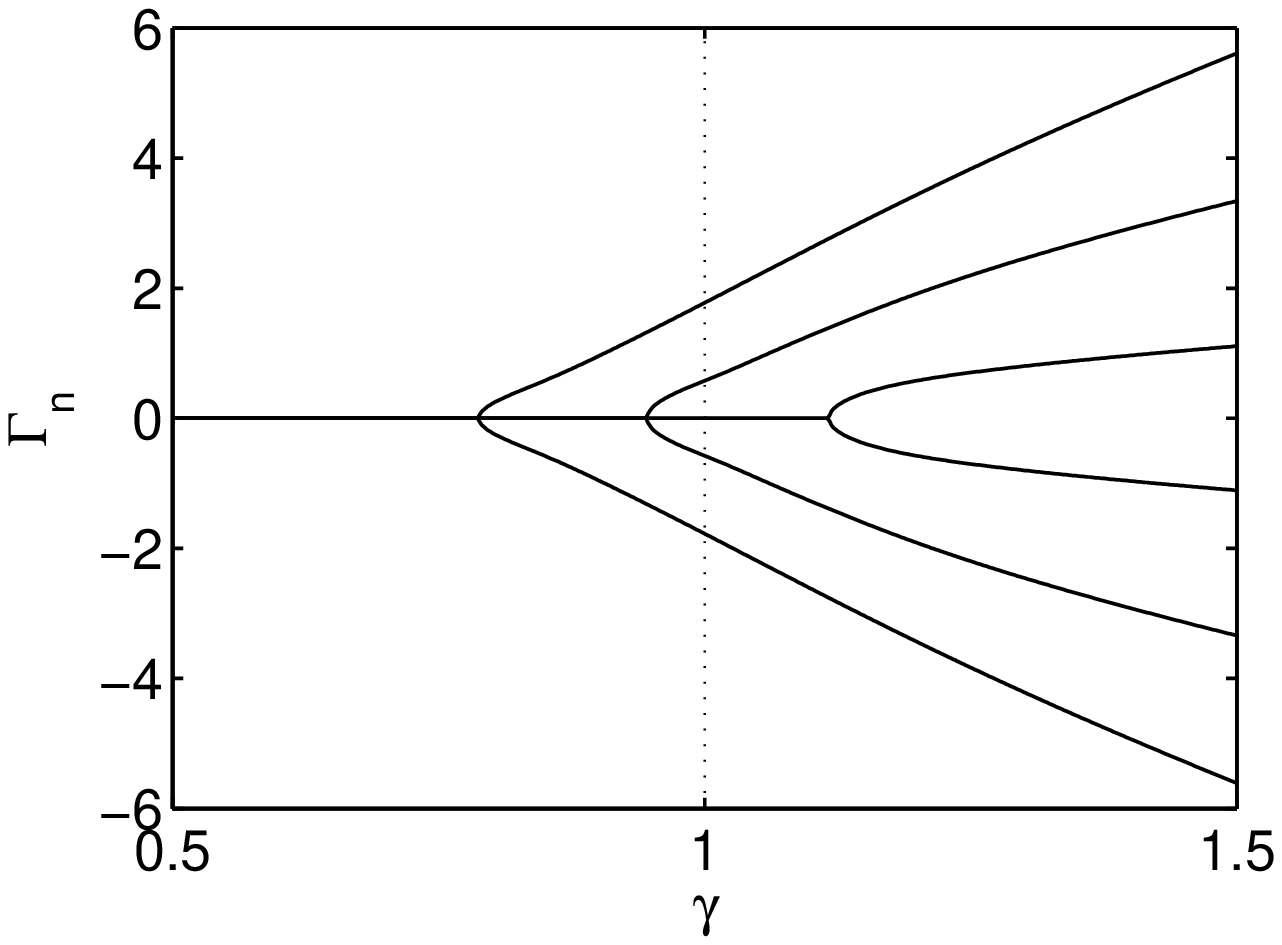}
\caption{\label{Spektrum6gross} Real- and imaginary parts of the
eigenvalues $\lambda_n=E_n-\rmi\Gamma_n$ of the Bose-Hubbard
Hamiltonian \rf{pt-ham} as a function of the non-Hermiticity
$\gamma$ for $v=1$, $N=5$ particles and $c=0.1/N$.}
\end{center}
\end{figure}

Figure \ref{Spektrum6gross} shows the real parts (left graphics) and
the imaginary parts (right graphics) of the numerically evaluated
eigenvalues of the Hamiltonian \rf{example1} for a small interaction
strength $c=0.1/N$. On the line $\gamma=v$ we find two pairs of
complex conjugate eigenvalues with positive real parts and two
purely real negative eigenvalues, in perfect agreement with the
qualitatively predictions of the first order perturbation
coefficients \rf{coeff_ex}.

\begin{figure}[!htb]
\begin{center}
  \includegraphics[width=8cm]{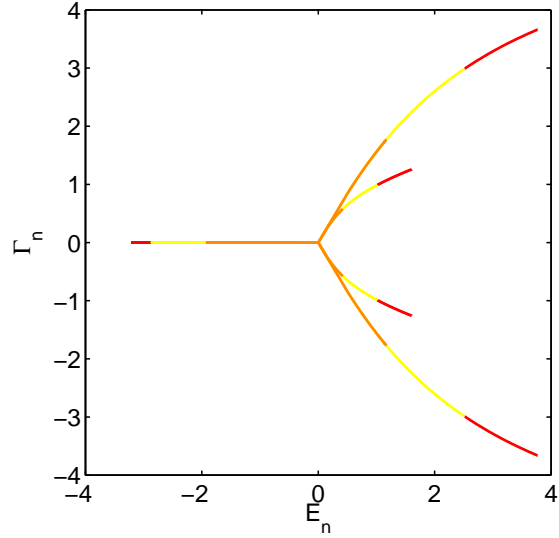}
\caption{\label{fig_EVtrajectory} (Color online) Trajectories of the
complex energy eigenvalues $\lambda_n=E_n-\rmi\Gamma_n$ of the
Bose-Hubbard Hamiltonian \rf{pt-ham} as a function of $c$ with
$0<cN<1$ (red), $0<cN<0.5$ (yellow) and $0<cN<0.1$ (orange), for
$\gamma=v$, $N=5$ particles.}
\end{center}
\end{figure}

\begin{figure}[!htb]
\begin{center}
 \includegraphics[width=6.4cm]{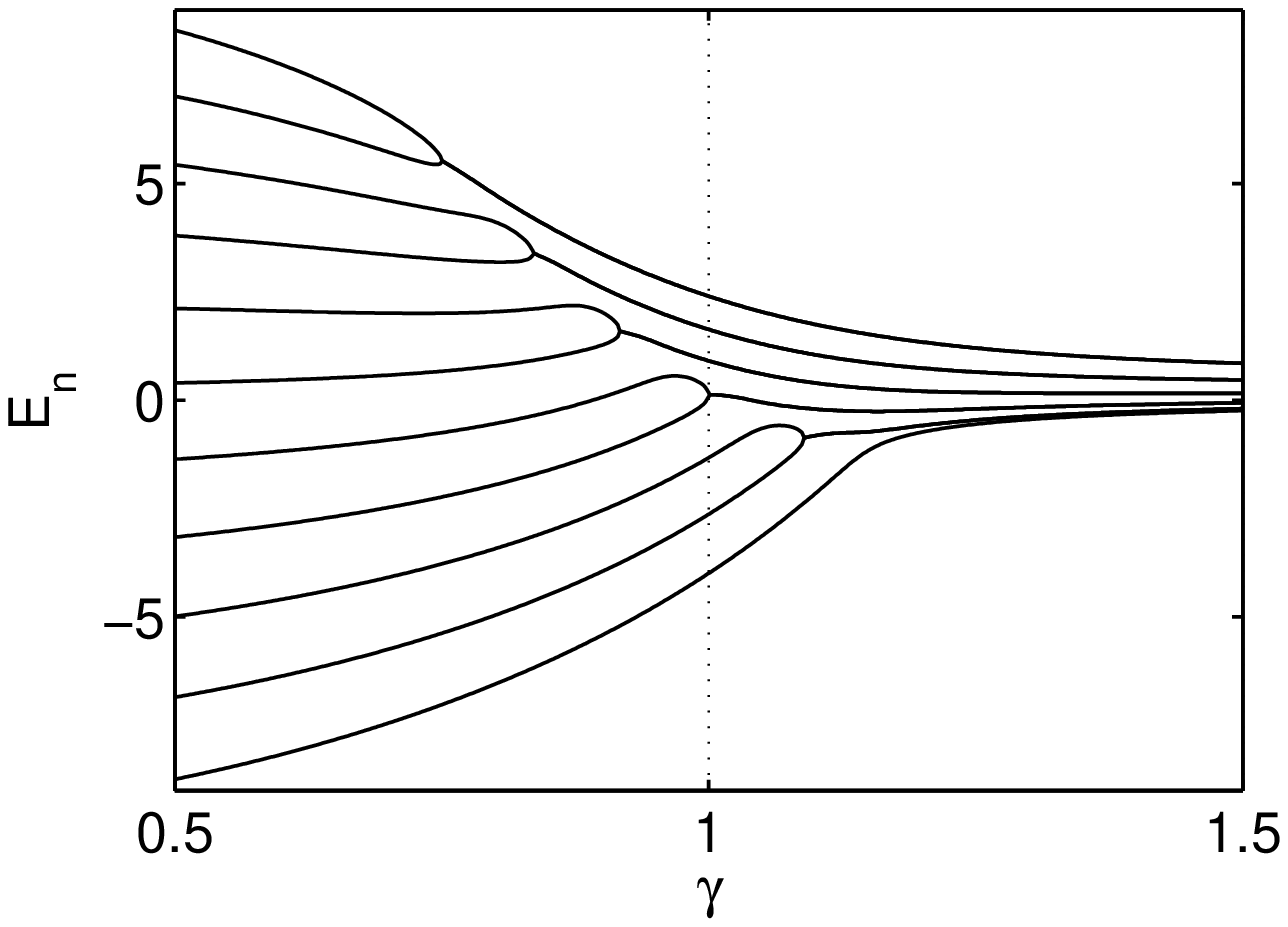}
 \includegraphics[width=6.4cm]{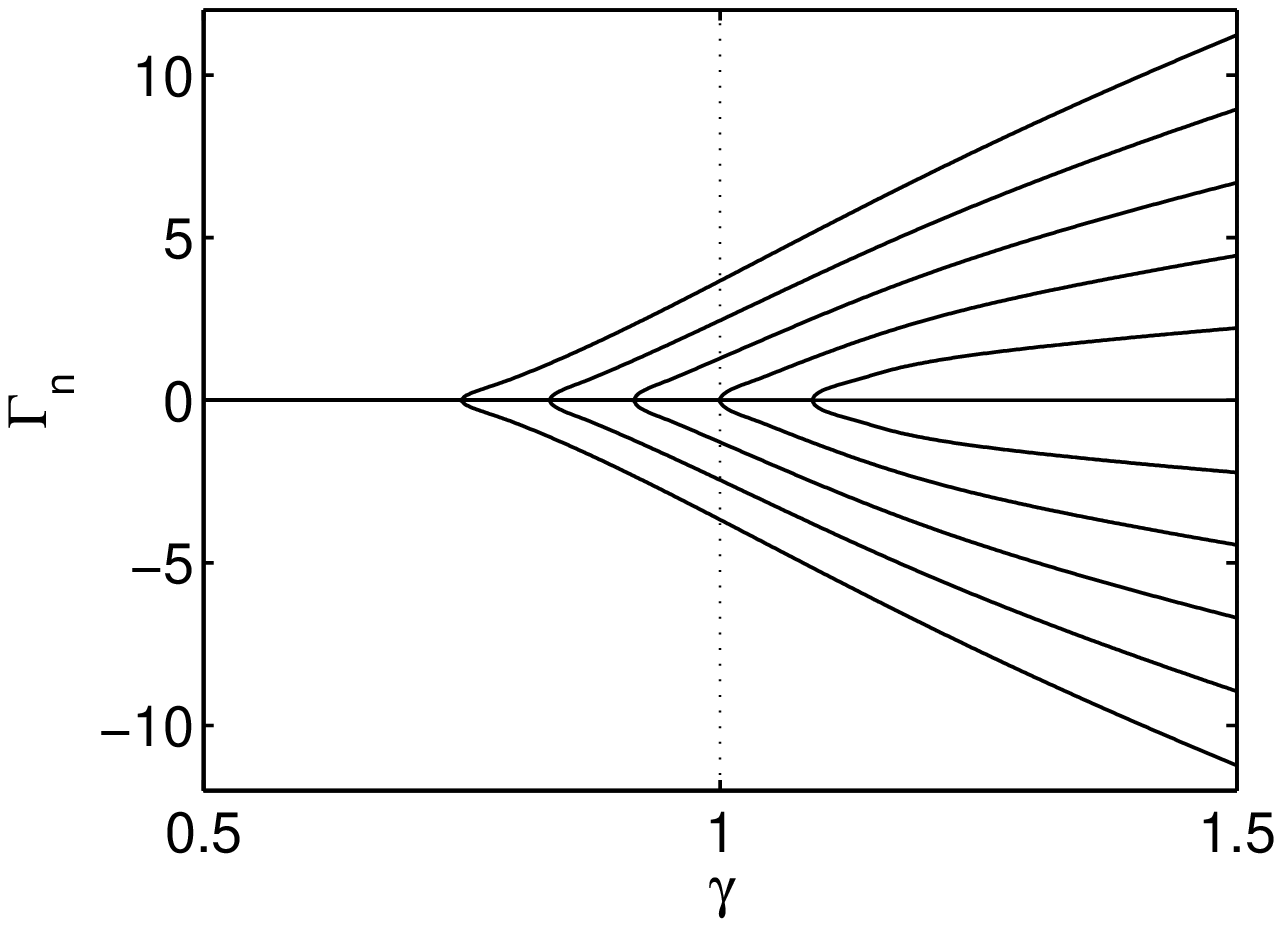}\\
 \includegraphics[width=6.4cm]{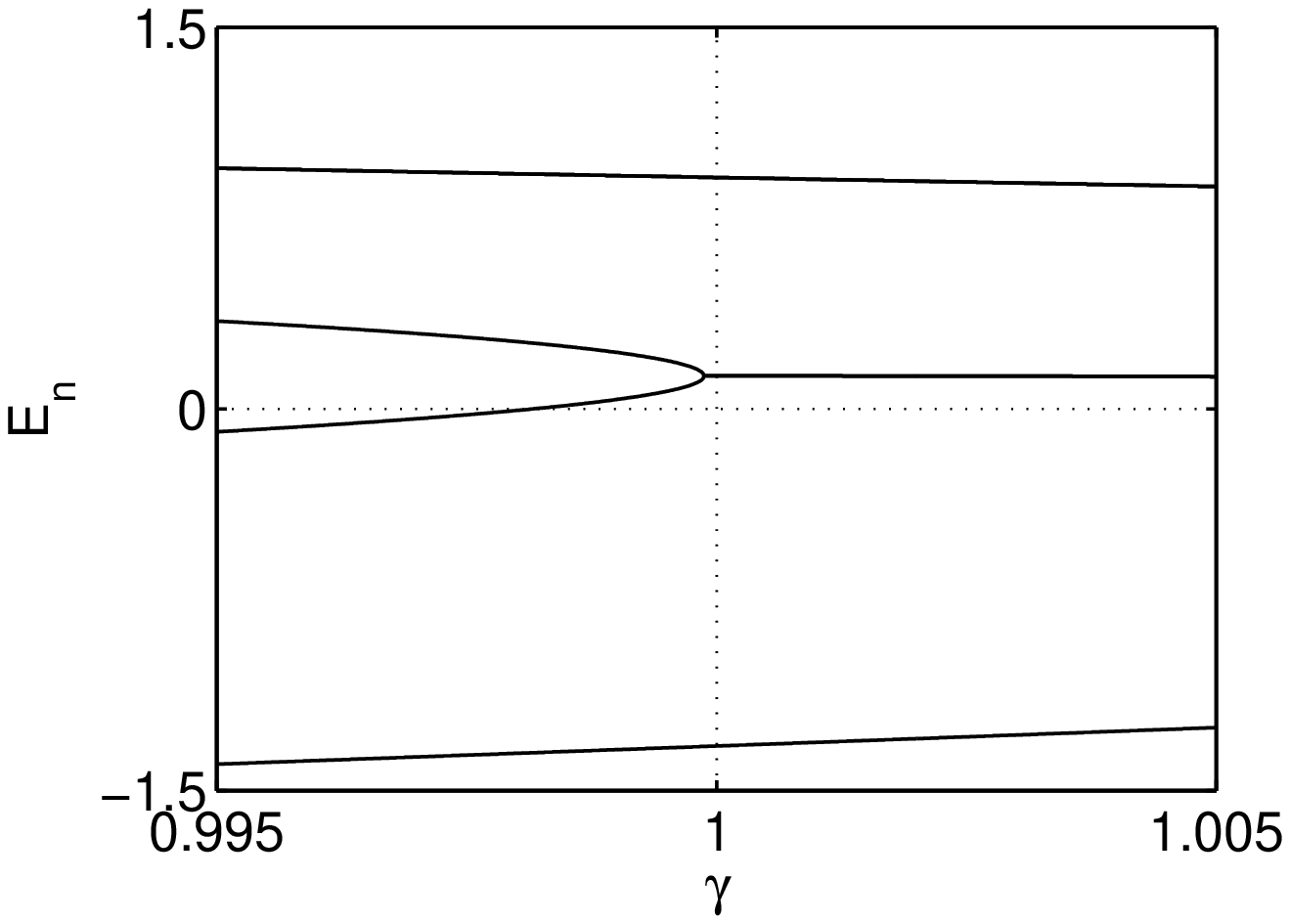}
\caption{\label{Spektrum7} Real- and imaginary parts of the
eigenvalues $\lambda_n=E_n-\rmi\Gamma_n$ of the Bose-Hubbard
Hamiltonian \rf{pt-ham} as a function of the non-Hermiticity
$\gamma$ for $v=1$, $N=10$ particles and $c=0.1/N$. The lower figure
shows a magnification of the real parts near the central EP.}
\end{center}
\end{figure}

\begin{figure}[!htb]
\begin{center}
  \includegraphics[width=6cm]{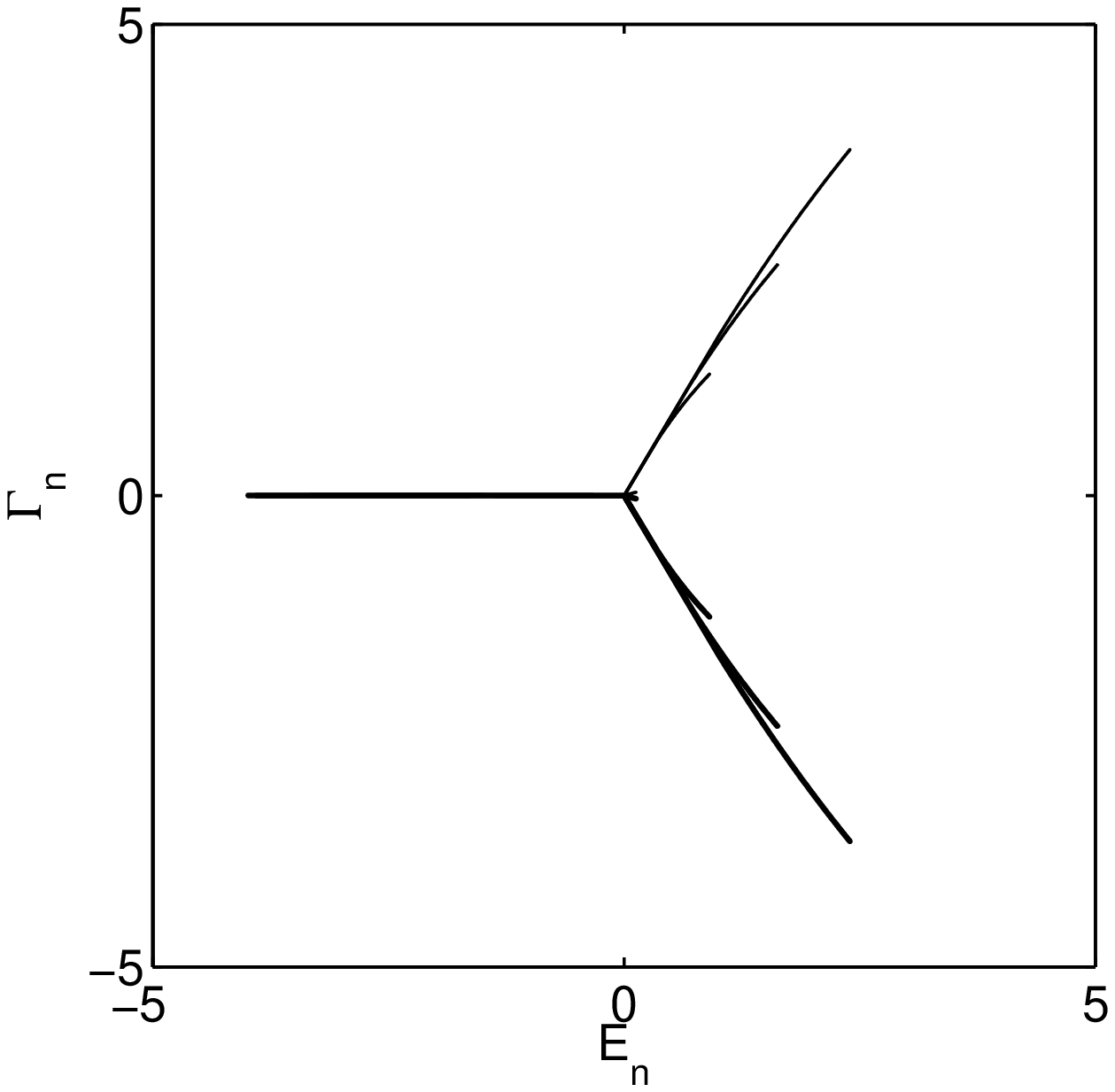}
  \includegraphics[width=6cm]{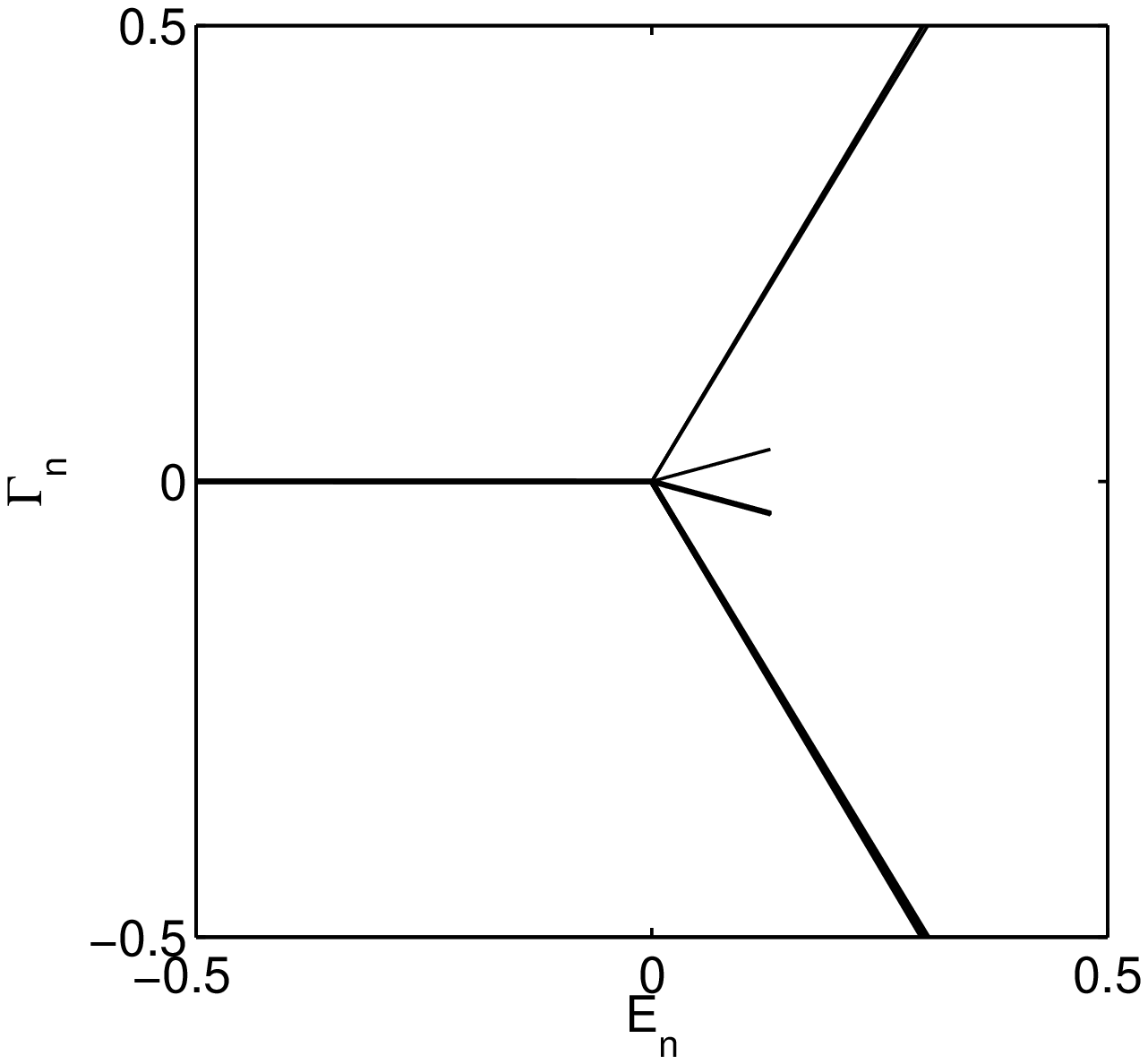}
\caption{\label{fig_EVtrajectory10} Trajectories of the complex
energy eigenvalues $\lambda_n=E_n-\rmi\Gamma_n$ of the Bose-Hubbard
Hamiltonian \rf{pt-ham} as a function of $c$ with $0<cN<0.1$, for
$\gamma=v$, $N=10$ particles and a magnification of the innermost
region (right).}
\end{center}
\end{figure}

The numerically obtained eigenvalue trajectories in the complex
plane for  $c\in (0,1)$ and fixed $\g=v$ are shown in
figure~\ref{fig_EVtrajectory}. For small $c\neq 0$ we find again the
typical $3-$rings. Their radii are given by the two different
absolute values of the coefficients \rf{coeff_ex}. For larger values
of $c$ the cubic-root-behavior becomes deformed by higher order
corrections what is clearly visible as a deviation from the straight
lines~$\sim e^{-i\pi \pm i\frac{2\pi}3}$. We note that the exact
trajectories can be approximated in terms of a series expansion with
arbitrary precision. Such series expansions are well defined over
parameter regions which lie within the convergence radius of the
series and break down when the next located algebraic singularity
(branch point, EP) is reached. (The distance to the next EP defines
the convergence radius (see, e.g., \cite{eom-singular}).)

Let us now turn to the model with $N=10$ particles. The
corresponding Puiseux-Newton-diagram in figure \ref{Puiseux} shows
again the typical modulo-three-ratchet-structure. The two straight
lines which form the lower boundary of the convex hull of the point
set have slopes $-1/3$ and $-1$  so that the $11$th-order EP unfolds
with dominant scaling powers $\mu_1^{(1)}=1/3$ and $\mu_1^{(2)}=1$,
i.e. as $\lb\sim c^{1/3}$ and $\lb\sim c$. Reinserting these
solutions into the corresponding characteristic polynomial yields
one ninth-order equation for the coefficient $e_1$ in $\lb\approx
e_1 c^{1/3}$. This ninth-order equation reduces to a cubic equation
in $e^3_1 $ and gives the three different values $|e_1|$ as scaling
parameters (ring radii) of the three triplets which comprise the
first nine coefficients $e_1^{(1)},\ldots, e_1^{(9)}$. The result is
similar to eq. \rf{coeff_ex} only with three triplets instead of
two. The quadratic equation for the remaining two coefficients
$e_1^{(10)}, e_1^{(11)}$ in the linear scaling law $\lb\approx e_1
c$ is easily derived by computer algebra and takes the explicit form
\be{eq_eN10}
-46423756800e_1^2+2410418995200e_1-33581039616000=0.
\ee
It yields the coefficients
\ba{eq_eN10_2}
e_1&=&\frac{145304}{5597}\pm\sqrt{\left(\frac{145304}{5597}\right)^2-\frac{4048640}{5597}}
\approx 26\pm 7\rmi .
\ea
Real and imaginary part of the spectrum as well as the unfolding of
the $11$th-order EP at $c=0$ and $\g=v=1$ are shown in figures
\ref{Spektrum7} and \ref{fig_EVtrajectory10}. Clearly visible in
figure~\ref{fig_EVtrajectory10} are the three $3-$rings and the two
linearly scaling single eigenvalues. Obviously, eigenvalue shifts
induced by the linear scaling are much smaller than the shifts of
the $3-$rings. In figures \ref{Spektrum7} and
\ref{fig_EVtrajectory10} these higher-order $(\lb\sim c)$
corrections are only visible in the zoomed graphics. In leading
$c^{1/3}-$order approximation, the EP related to the $\lb\sim c$
branches remained fixed at the original position $\g=v=1$.

Summarizing we conclude that for arbitrary particle number $N$ the
Newton-diagram at the $(N+1)$th-order EP shows a
modulo-three-ratchet-structure with regard to the unfolding due to
increasing interaction strength $c$ --- like in figure
\ref{Puiseux}. For non-vanishing  $c$ the EP unfolds into
$\left[\frac{N+1}{3}\right]$ eigenvalue triplets forming regular
$3-$rings in the complex plane with dominant scaling behavior of the
type $c^{1/3}$ for $|c|\ll |v|/N$. The remaining $(N+1)\mod3$ single
eigenvalues depend linearly on $c$. Furthermore we find for small
interaction strength $c$ and an original $(N+1)$th-order EP at
$|\g|=|v|$ that roughly $2/3$ of the occurring second-order EPs are
located in parameter regions $0<|\gamma|<|v|$  and roughly $1/3$ in
the region $|v|<|\g|$ what confirms the numerical results of section
\ref{sec_num}.

It remains to emphasize that the unfolding of $(N+1)$th-order EPs
into triplets (for $\g^2=v^2$) has its origin in the effective
$3-$Hessenberg form of the perturbation matrix \rf{h-2}, \rf{h-3}.
This suggests to reinterpret the square-root spectrum of the exactly
solvable $(c=0)-$model from section \ref{sec_lim_int} as EP
unfolding under perturbation by a $2-$Hessenberg matrix. Indeed,
representing the Hamiltonian \rf{c0-1} as
\be{h30}
H=2v(L_x-iL_z)-2i\Delta L_z\qquad \qquad \Delta:=\g-v
\ee
and performing an $SU(2)$ rotation as for \rf{h-2} it takes the
structure
\be{h31}
\tilde H=2vL_--\Delta (L_+-L_-).
\ee
Due to the fact that $L_-$ is of Jordan block type the
Hessenberg perturbation theory of \cite{ma} is applicable.
The perturbation matrix $\Delta (L_+-L_-)$ has non-vanishing
entries only on the first sub- and superdiagonals and it is
therefore of $2-$Hessenberg type. According to \cite{ma} the
$(N+1)$th-order EP at $\Delta=0$ unfolds then under this
$2-$Hessenberg type perturbation into $[\frac{N+1}2]$ eigenvalue
pairs and,  for $N$ even, into one additional single eigenvalue.
Obviously, this prediction is in complete agreement with the exact
result \rf{Ev_lin} for the spectrum which shows a square root
(pairwise) unfolding of the mother EP and an additional single
eigenvalue $\lb=0$ in case of even $N$.

The EP unfolding according to the Hessenberg perturbation type is
straightforwardly extendable to Hamiltonians with $\g=v$ and
higher-order perturbations in $L_z$
\be{h32}
H=2v(L_x-iL_z)+c L_z^k\qquad \qquad k\ge 2.
\ee
In this case an  $SU(2)$ rotation leads to
\be{h33}
\tilde H=2vL_-+ \left(-\frac i2\right)^k c  (L_+-L_-)^k.
\ee

\begin{figure}[!htb]
\begin{center}
\includegraphics[width=6cm]{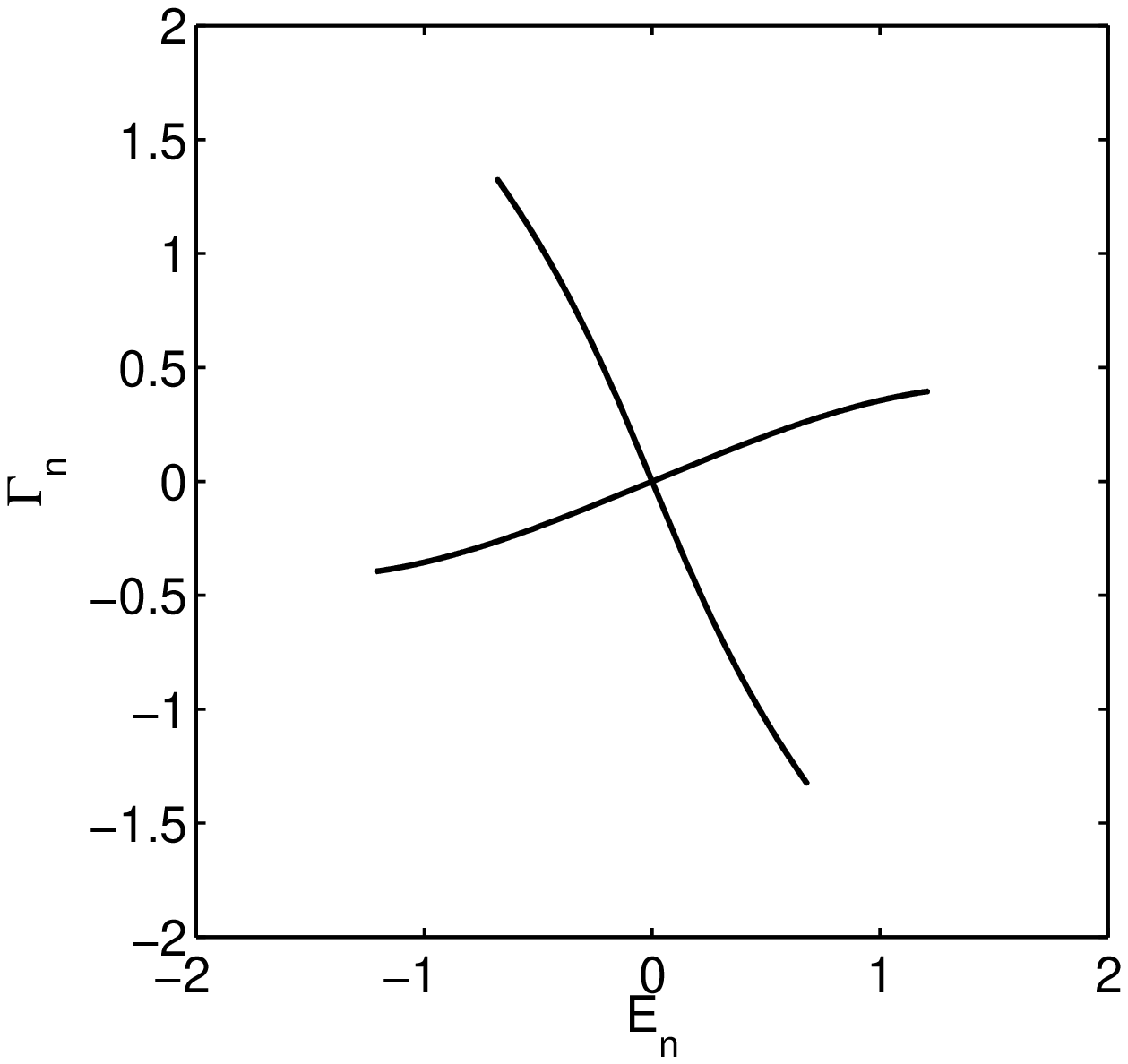}
\includegraphics[width=6cm]{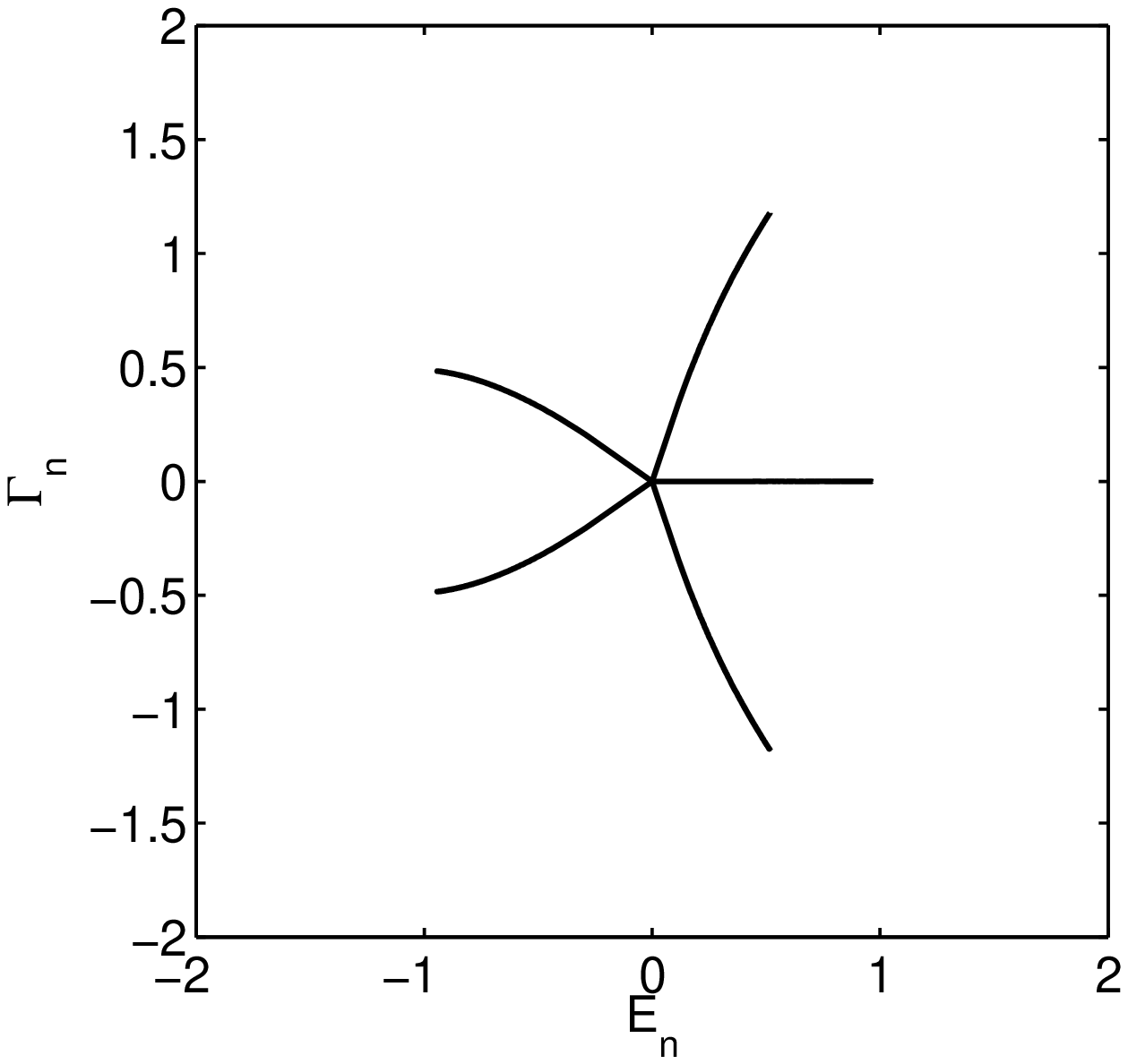}
\caption{\label{fig_unfolding45} Eigenvalue trajectories of a
$(N=4)-$particle Bose-Hubbard Hamiltonian \rf{h32} with fixed
$v=1$ for $k=3$ and $0<c<0.1/N^2$ (left, $4-$ring) and for $k=4$ and
$0<c<0.1/N^3$ (right, $5-$ring).}
\end{center}
\end{figure}

For $1\le k\le N$ the perturbation matrix $(L_+-L_-)^k$ is of
$(k+1)-$Hessenberg type so that the $(N+1)$th-order EP will unfold
into $[\frac{N+1}{k+1}]$ eigenvalue rings of size $k+1$ and $r=
(N+1) \mod (k+1)$ eigenvalues which will be grouped in one or
several smaller rings. Figure \ref{fig_unfolding45} shows the EP
unfolding for  $(N=4)-$particle Hamiltonians \rf{h32} with $k=3$ and
$k=4$, small $c$ and fixed $\g=v=1$. Clearly visible are the
$4-$ring and $5-$ring eigenvalue structures. For $k\ge N$ also the
lowest left matrix element  becomes, in general, nonvanishing
$H_{N+1,1}\neq 0$ so that for these $k$ the mother EP will unfold
into a single $(N+1)-$ring of eigenvalues.

\subsection{The limit of strong interaction\label{strong coupling}}
To understand the limit of strong interaction analytically, we can
apply ordinary perturbation theory with $\gamma$ and $v$ being the
small parameters:
\begin{equation}
H = \underbrace{2 c  L_z^2}_{H_0} \underbrace{-2 {\rm i}\gamma L_z +
2 v  L_x}_{H_1}.
\end{equation}
In general the eigenstates of $H_0$ are doubly degenerate,
which can be seen in the standard basis \rf{ang-mom-4}
\begin{equation}
 2 c  L_z^2 |l,\pm m_z \rangle = 2 c m_z^2 |l, \pm m_z \rangle,
\end{equation}
with $l=\frac{N}{2}$. An exclusion is the eigenstate  with $m_z=0$
in a system with even particle number $N$. This state is not
degenerate.

In lowest-order approximation, the perturbed degenerate energy
levels are given as
\begin{equation}
E \simeq E_0+E_1\,,
\end{equation}
where the corrections $E_1$ can be calculated from the perturbation
matrix
\begin{equation}
 W=\langle l, m_z | H_1 |l,  m_z' \rangle \qquad \qquad m_z,m_z'=\pm |m_z|.
\end{equation}
We start with even particle numbers $N$. In this case the matrix $W$
is diagonal and the energy corrections read
\begin{equation}\label{correction1}
 E_1=- 2 {\rm i} \gamma m_z.
\end{equation}
Obviously, the state $m_z=0$ remains unperturbed in lowest order
approximation.

For odd particle numbers $N$ the correction (\ref{correction1})
holds as well, except in the case of $|m_z|=1/2$ where the
perturbation matrix is nondiagonal
\begin{equation}
 W=\left(
\begin{array}{cc}
 - {\rm i}\gamma & v \frac{N+1}2\\
 v \frac{N+1}2 & {\rm i}\gamma
\end{array}
\right).
\end{equation}
The eigenvalues of this matrix are
\begin{equation}\label{correction2}
 E_1=\pm \sqrt{v^2 \left(\frac{N+1}{2}\right)^2 - \gamma^2}
\end{equation}
and we see that there occur two second-order EPs at $\g=\pm
\frac{N+1}2 v$.  The corrections $E_1$ for the states
$|m_z|=\frac12$ are purely real for $|\gamma| < |v| \frac{N+1}{2}$
and purely imaginary for $|\gamma| > |v| \frac{N+1}{2}$.

The exact spectrum \rf{Ev_lin} for $c=0$ and the numerical studies
for $c\neq 0$ show that pairwise complex conjugate eigenvalues of
$H$ occur for large values of $|\g|$. In connection with the purely
imaginary perturbative corrections $E_1$ for states with $|m_z|>1/2$
this implies that for large $|c|\gg |v|/N$ all EPs involving these
states must have tended to $\g\to 0$.

Summarizing we conclude that in the limit $c\to \infty$ there exists
one zero-eigenvalue state with $m_z=0$ for $N$ even and a pair of
real eigenvalues for states $|m_z|=1/2$ in the parameter region
$|\gamma| < |v| \frac{N+1}2$ of a model with $N$ odd. All remaining
eigenvalues come as complex conjugate pairs.

Therefore these perturbative results prove the numerical
observations of section \ref{sec_num}, that in the limit of
ultra-strong interaction all EPs of an $N-$even model are located
at $\gamma=0$, whereas in an $N-$odd model two of the EPs can  be
found at $\gamma=\pm v\frac{N+1}{2}$.

\section{Conclusion and Outlook}
We studied the spectrum of a non-Hermitian $\cP\cT-$symmetric
two-mode Bose-Hubbard Hamiltonian, a system modeling an $N-$particle
Bose-Einstein condensate in a double well potential containing a
sink in one of the wells and a source of equal strength in the
other. While for vanishing particle interaction there exists only
one pair of EPs of order $N+1$, the interplay of non-Hermiticity and
particle interaction leads to a characteristic unfolding of these
EPs into $3-$rings of eigenvalues and the occurrence of a series of
EPs of order two. This numerically observed scenario has been
analytically understood using the Puiseux-Newton perturbation
technique. Furthermore the case of strong particle interaction was
described by ordinary Rayleigh-Schr\"odinger perturbation theory.

Further investigations concerning, e.g., the positions of the EPs as
well as their influence on the system dynamics remain  tasks for
future research.

Another challenge is the investigation of the large $N$ limit of the
present model, resp. the so called mean-field approximation.
In the Hermitian case this mean-field approximation is usually achieved by
replacing the bosonic field operators by c-numbers,
the condensate wave functions, yielding the nonlinear
Schr\"odinger equation resp. Gross-Pitaevskii equation.
This approach is closely related to a classicalization. In a
number of recent papers consequences of the classical nature of the
mean-field approximation are discussed and semiclassical aspects are
introduced \cite{Vard01b, Angl01, Mahm05, Moss06, 06zener_bec,Wu06}.
For a two-mode system even the eigenenergies and eigenstates of the
many particle system could be reconstructed approximately from the
mean-field system in a semiclassical approximation with astonishing
accuracy \cite{semi07}. While there are some investigations
concerning an heuristically introduced non-Hermitian generalization
of discrete nonlinear Schr\"odinger equations \cite{Schl04,
Hill06,06nlnh,06nlres,Livi06,Fran07} a careful derivation of a
mean-field approximation starting from a non-Hermitian many particle
system was lacking in the past and will be the subject of a separate
paper \cite{08nhbh}.

\section*{Acknowledgments}
We thank Oleg Kirillov for useful comments on \cite{oleg2004} and on
Lidski's technique mentioned, e.g., in \cite{ma}. Support from the
Deutsche Forschungsgemeinschaft via the Graduiertenkolleg
``Nichtlineare Optik und Ultrakurzzeitphysik'' and the Collaborative
Research Center SFB 609 is gratefully acknowledged.

\end{document}